\documentclass[aps,twocolumn,preprintnumbers,floats]{revtex4}\def\@cite#1#2{\textsuperscript{[{#1\if@tempswa , #2\fi}]}}

\usepackage{graphicx}
\usepackage{amsmath}
\usepackage{amsfonts}
\usepackage{amssymb}
\usepackage{color}
\usepackage{subfigure}
\usepackage{epsfig}
\usepackage{epstopdf}
\usepackage{morefloats}
\usepackage{multirow}
\usepackage{graphicx,booktabs}
\usepackage{mathrsfs}
\usepackage{txfonts}
\usepackage{indentfirst}
\usepackage{graphicx,booktabs}
\usepackage{longtable,lscape}
\usepackage[figuresright]{rotating}
\usepackage[colorlinks, citecolor=blue,anchorcolor=red,menucolor=red, linkcolor=red,filecolor=red,urlcolor=blue,frenchlinks=red]{hyperref}

\newcommand{\vsig}{\mbox{\boldmath$\sigma$\unboldmath}}

\begin{document}
	
\title{Unified unquenched quark model for heavy-light mesons with chiral dynamics}

\author{Ru-Hui Ni$^{1}$,
Jia-Jun Wu$^{2,3}$~\footnote{E-mail: wujiajun@ucas.ac.cn},
and
Xian-Hui Zhong$^{1,4}$~\footnote{E-mail: zhongxh@hunnu.edu.cn}}

\affiliation{$^{1}$Department of Physics, Hunan Normal University, and Key Laboratory of Low-Dimensional Quantum Structures and Quantum Control of Ministry of Education, Changsha 410081, China}

\affiliation{$^{2}$School of Physical Sciences, University of Chinese Academy of Sciences (UCAS), Beijing 100049, China }
\affiliation{$^{3}$Southern Center for Nuclear-Science Theory (SCNT), Institute of Modern Physics, Chinese Academy of Sciences, Huizhou 516000, Guangdong Province, China }

\affiliation{$^{4}$Synergetic Innovation Center for Quantum Effects and Applications (SICQEA), Hunan Normal University, Changsha 410081, China}

\begin{abstract}
In this work, an unquenched quark model is proposed for describing the heavy-light mesons by taking into account the coupled-channel effects induced by chiral dynamics. After including a relativistic correction term for the strong transition amplitudes, both the mass spectra and
decay widths of the observed heavy-light mesons can be successfully described simultaneously in a unified framework, 
several long-standing puzzles related to the small masses and broad widths are overcome naturally.
We also provide valuable guidance in searching new heavy-light mesons by the detailed predictions of their masses, widths, and branching ratios.
The success of the unquenched quark model presented in this work indicates
it may be an important step for understanding the hadron spectrum.

\end{abstract}

\maketitle

\section{Introduction}
The fundamental theory of strong interaction is Quantum Chromodynamics (QCD).
One of its most prominent challenges is the phenomenon of confinement, the elementary systems we observe, known as hadrons and composed of quarks and gluons, appear to be colorless.
In the low-energy regime, the non-perturbative effects make it impossible to achieve analytical computations.
Lattice QCD simulation can provide the hadronic spectra from first principles, but it falls short of offering a detailed picture of hadrons.
In such circumstances, a theoretical model becomes essential for gaining a deeper insight into the nature of confinement.

The quenched quark model, initially based solely on $q\bar{q}$ for mesons and $qqq$ for baryons and developed in the 1960s by Gell-Mann and Zweig~\cite{Gell-Mann:1964ewy,Zweig:1964ruk,Zweig:1964jf}, effectively described the hadron spectrum until the discovery of the $X(3872)$~\cite{Belle:2003nnu} and $D_s(2317)$ in 2003~\cite{BaBar:2003oey}.
Over the past two decades, an increasing number of hadron states, often referred to as `exotic', have been observed~\cite{ParticleDataGroup:2022pth}.
These discoveries, including the $XYZ$ and $P_c$ states, have challenged the predictions of the simple $q\bar{q}$
and $qqq$ models~\cite{Deng:2016ktl,Deng:2016stx,Ebert:1997nk,Liu:2015lka,Capstick:1986ter,Eichten:1978tg,Godfrey:1985xj,Swanson:2005,Godfrey:2015dva,Godfrey:2016nwn,Ebert:2009ua, Zeng:1994vj,Liu:2016efm,Liu:2013maa,Vijande:2004he,Li:2010vx,Lu:2016bbk,Asghar:2018tha,Colangelo:2012xi,Sun:2014wea,Song:2015fha,Song:2015nia,Ferretti:2015rsa,DiPierro:2001dwf,Zhong:2008kd,Zhong:2010vq,Zhong:2009sk,Xiao:2014ura,Ni:2021pce}, more works in the reviews~\cite{Capstick:2000qj,Chen:2022asf,Chen:2016spr}.
An undeniable factor contributing to this discrepancy is the omission of coupled-channel effects resulting from hadron loops in these quark models.
This deficiency has long been recognized and discussed, for instance, in 1980, a similar effect known as the meson cloud was proposed~\cite{Theberge:1980ye}.
This underscores the importance of considering interactions at both the quark(gluon) level and the hadron level to obtain a comprehensive understanding of physical hadrons.
	
The significance of coupled-channel effects is widely acknowledged within the scientific community.
However, conducting a systematic study in practical calculations presents several challenging issues.
The primary concerns can be summarized as follows:
	\begin{itemize}
		\item[{\bf P1.}] How to select the coupled-channels.
		\item[{\bf P2.}] How to obtain correct both mass and width.
		\item[{\bf P3.}] How to evaluate the coupled channel effects in the high momentum region.
	\end{itemize}
	
All of these challenges add complexity when employing a comprehensive model.
Consequently, existing research has primarily focused on exploring coupled-channel effects for states that deviate significantly from conventional quark models, rather than those with well-established explanations within the hadron spectrum.
For instance, while considering loop contributions for scalar mesons, such as $K\bar{K}$ and $\pi\pi$ loops, only quark level interactions are accounted for $\rho$ meson where the $\pi\pi$ loop indeed significantly shifts the mass of $\rho$.
This lack of consistency and self-consistency within the theory arises as an issue.
Furthermore, in the context of strong decay, systematic studies have been conducted within several phenomenological models~\cite{Godfrey:2015dva,Godfrey:2016nwn,Asghar:2018tha,Li:2010vx,Lu:2016bbk,Ferretti:2015rsa,Song:2015nia,Sun:2014wea,Song:2015fha,Colangelo:2012xi,DiPierro:2001dwf,Zhong:2008kd,Zhong:2010vq,Zhong:2009sk,Xiao:2014ura,Ni:2021pce}, as detailed in recent reviews~\cite{Chen:2022asf,Chen:2016spr}.
However, these models often struggle to accurately describe the widths of higher resonances~\cite{Chen:2022asf,Chen:2016spr}.

%
In the past twenty years, numerous excited heavy-light meson candidates have been observed by collaborations such as D0, CDF, \emph{BABAR}, and LHCb~\cite{ParticleDataGroup:2022pth}.
Most theoretical studies have primarily focused on near-threshold states, such as $D_{s0}^*(2317)$ and $D_{s1}(2460)$~\cite{Hwang:2004cd, Dai:2003yg, Zhou:2020moj, Wu:2011yb, Mohler:2013rwa, Lang:2014yfa}, or have been limited to the $D_s$ spectrum or $D$ spectrum separately~\cite{Zhou:2011sp,Xie:2021dwe,Hao:2022vwt}.
In comparison to the mass spectrum, $D_0(2550)$/$D_{s0}(2590)$ and $D^*_1(2600)$/$D^*_{s1}(2700)$ can be categorized as radial excitations.
However, the predicted decay widths within the chiral quark model~\cite{DiPierro:2001dwf,Zhong:2008kd,Ni:2021pce,Xiao:2014ura,Zhong:2010vq,Zhong:2009sk} systematically underestimate the observed values.
Thus, a comprehensive investigation of both the mass spectra and decay widths for all heavy-light mesons, including $D$, $D_s$, $B$, and $B_s$ mesons, within a unified framework that incorporates unquenched coupled-channel effects is currently lacking.
This is a crucial step in the development of a comprehensive model for hadron physics.

In this work, we will fill this gap by introducing a unified unquenched quark model framework that accounts for coupled-channel effects, utilizing
a semi-relativistic potential to interpolate both the masses and widths of all heavy-light mesons for the first time.

\begin{figure*}
\centering \epsfxsize=16.0 cm \epsfbox{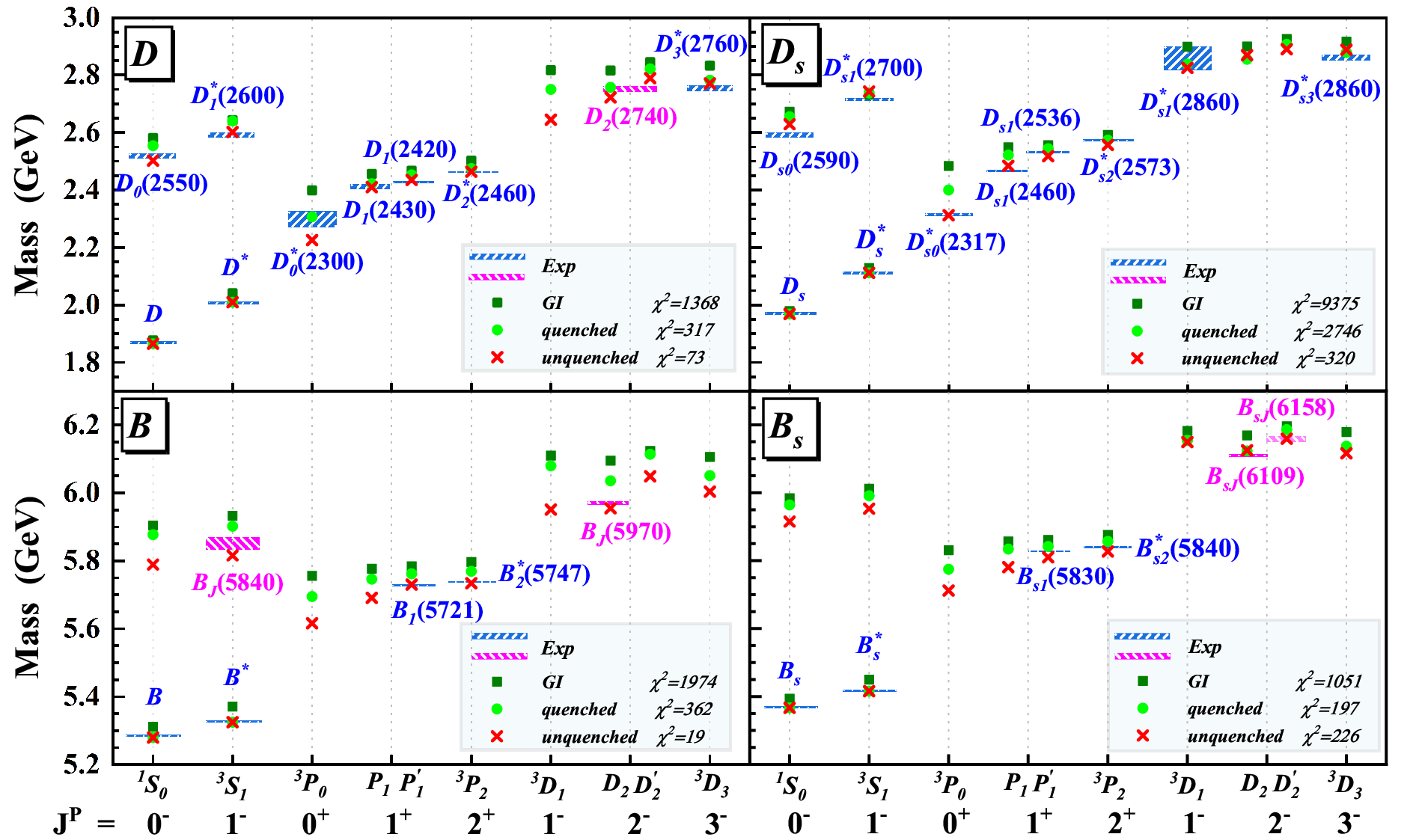} \vspace{-0.2 cm} \caption{Mass spectra of heavy-light mesons compared with the observations.
The $\chi^2=\sum_{i}(\mathrm{Thy}(i)-\mathrm{Exp}(i))^2/\mathrm{Error}(i)^2$ for the well-known GI model, our quenched model and unquenched spectrum, respectively.Here $\mathrm{Thy}$, $\mathrm{Exp}$ and $\mathrm{Error}$ represent the theoretical results, experimental data and error, respectively.
It should be noted that for data with experimental errors less than $1$ MeV, we unified set $\mathrm{Error}=2$ MeV because our systemic error should be larger than several MeV. }\label{Mspectrum}
\end{figure*}

\section{Framework}

%
In this model, as described in previous works~\cite{Heikkila:1983wd,Kalashnikova:2005ui,Barnes:2007xu,Lu:2016mbb, Ortega:2009hj}, the Hamiltonian of the hadronic system is described by
\begin{equation}
		\mathcal{H} = \mathcal{H}_0+\mathcal{H}_c +\mathcal{H}_I.
\end{equation}
Here, $\mathcal{H}_0$ represents the Hamiltonian governing
the bare $Q\bar{q}$ state denoted as $|A\rangle$, and it is derived from the semi-relativistic quark potential model.
$\mathcal{H}_c$ is the noninteracting Hamiltonian for the
continuum state $|BC\rangle$,
while $\mathcal{H}_I$ is an effective Hamiltonian for
describing the coupling between $|A\rangle$ and $|BC\rangle$.
The quark (gluon) level interactions are all encompassed within $\mathcal{H}_0$, which are clearly defined, as introduced in the Supplemental Material~\cite{SuppMat}.
Meanwhile, $\mathcal{H}_I$ contains the hadronic interactions, which will be explored in detail.
	
To address the question posed in {\bf P1} on the first page, the mass of a dressed hadron
is estimated with the once-subtracted method suggested in Ref.~\cite{Pennington:2007xr}, i.e.,
	\begin{equation}\label{Mass shift}
		\begin{aligned}
			M= M_A-\operatorname{Re} \sum_{BC} \int_{0}^{\infty}
			 \frac{\left(M_0-M\right)\overline{\left|\langle BC,\boldsymbol{q}|\mathcal{H}_{I} |A \rangle
					 \right|^{2}}}{\left(M-E_{BC}\right)\left(M_0-E_{BC}\right)}
			q^{2} d q,
		\end{aligned}
	\end{equation}
where $\boldsymbol{q}\equiv(0,0,q)$, and $M_{A}$ is the bare mass determined by the $\mathcal{H}_0$.
$M_{0}$ represents the subtracted zero point for the heavy-light meson system. $E_{BC}$ is the kinematic energy of the $|BC,\boldsymbol{q}\rangle$ continuum state with the loop momentum $\boldsymbol{q}$.
We select the masses of the ground heavy-light meson states, namely $D_{(s)}$ and $B_{(s)}$, as the reference values denoted by $M_{0}$.
With the once-subtracted method, in the calculations one only need consider the OZI-allowed two-body hadronic channels
with mass thresholds below or just above the bare $|A\rangle$ states. The contributions from the other virtual channels
(whose mass thresholds far above the bare $|A\rangle$ states) are
subtracted from the dispersion relation by redefining the bare mass.
This method has been applied to study the coupled-channel effects
on $D/D_s$ meson states and charmonium states in the literature~\cite{Zhou:2011sp,Duan:2021alw}.
It should be emphasized that the application of once-subtracted
method is a crucial step for obtaining a successful description of
the heavy-light meson spectrum within a unified unquenched quark model.

For a strong decay process $A\to BC$, the partial decay width is obtained by
\begin{eqnarray}\label{width}
		\Gamma=2\pi \frac{|\boldsymbol{q}| E_{B}E_{C}}{M}
		\overline{\left|\langle BC, \boldsymbol{q}|\mathcal{H}_{I} |A \rangle\right|^{2}},
\end{eqnarray}
where $\boldsymbol{q}$ and $E_{B/C}$ are the on-shell momentum and energy of particle $B/C$ in the decay, respectively.

The central challenge lies in the evaluation of the strong transition amplitude
$\langle BC,\boldsymbol{q} |\mathcal{H}_{I} |A \rangle$.
This amplitude can be determined using the chiral quark model incorporating chiral dynamics, as outlined in previous works~\cite{Manohar:1983md,Li:1994cy,Li:1997gd,Zhao:1998fn, Zhao:2000tb,Zhao:2001jw,Zhao:2002id,DiPierro:2001dwf,Zhong:2008kd,Zhong:2007gp,Zhong:2010vq,Zhong:2009sk,Xiao:2014ura,li:2021hss,Ni:2021pce}.

In the chiral quark model, the low-energy quark-antiquark-pseudoscalar-meson interaction is described by the chiral Lagrangian:~\cite{Li:1994cy,Li:1997gd,Zhao:2002id}:
\begin{equation}\label{PLg} {\cal L}_{P}=\sum_j\frac{1}{f_m}\bar{\psi}_j\gamma^{j}_{\mu}\gamma^{j}_{5}\psi_j\vec{\tau} \cdot\partial^{\mu}\vec{\phi}_m.
	\end{equation}
Here, $\psi_j$ represents the $j$-th quark field in the hadron,
$\phi_m$ is the pseudoscalar meson field, $\tau$ is an isospin operator, and $f_m$ is the pseudoscalar meson decay constant.

By carrying out a nonrelativistic expansion of the chiral Lagrangian up to the mass order of $1/m^2$,
one can obtain
\begin{equation}\label{Mass shift}
\mathcal{H}_{\mathcal{I}}=\mathcal{H}_{\mathcal{I}}^{NR}+\mathcal{H}_{\mathcal{I}}^{RC},
\end{equation}
where $\mathcal{H}_{I}^{NR}$ is the nonrelativistic term
at the order of $1/m$:
\begin{equation}\label{non-rela}
\begin{aligned}
&
{\cal H}_{\mathcal{I}}^{NR}=
g\sum_{j}\left[\mathcal{G}\boldsymbol{\sigma}_{j} \cdot \boldsymbol{q}+\frac{\omega_m}{2\mu_q}
(\boldsymbol{\sigma}_{j} \cdot \boldsymbol{p}_{j})\right]I_{j} \varphi_m,
\end{aligned}
\end{equation}
while $\mathbf{{\cal H}}_{I}^{RC}$ is the higher order term at the higher mass order of $1/m^2$, as a relativistic correction term of Eq.~(\ref{non-rela}), which is given by
\begin{eqnarray}
\mathcal{H}_{I}^{RC}&=&-\frac{g}{32\mu_{q}^2}\sum_{j}[m_{\mathbb{P}}^2(\boldsymbol{\sigma}_{j} \cdot \boldsymbol{q})\nonumber\\
&&+2\boldsymbol{\sigma}_{j}\cdot (\boldsymbol{q}-2\boldsymbol{p}_{j}) \times (\boldsymbol{q} \times \boldsymbol{p}_{j}) ]I_{j} \varphi_m.
\end{eqnarray}
In the above equations, $\boldsymbol{p}_{j}$ and $\vsig_j$ are the internal momentum operator and the spin operator
of the $j$-th light quark in a hadron, respectively.
$\varphi_m=e^{-i\boldsymbol{q} \cdot \boldsymbol{r}_{j}}$ is the plane wave part of the emitted light meson
with three-vector momentum and energy denoted by ($\boldsymbol{q}$, $\omega_m$). $m_{\mathbb{P}}$
stands for the mass of the light pseudoscalar meson.
$I_j$ is an isospin operator defined in the SU(3) flavor space~\cite{Li:1997gd}. The factors $g$ and $\mathcal{G}$ are defined by $g\equiv\delta \sqrt{(E_i+M_i)(E_f+M_f)}/f_m$ and $\mathcal{G}\equiv
-\left(1+\frac{\omega_{m}}{E_f+M_f}+\frac{\omega_{m}}{2m'_j}\right)$, respectively, where $\delta=0.557$ as a
dimensionless strength parameter is determined by our previous works~\cite{Zhong:2007gp,Zhong:2008kd}, while the decay constants $f_m$ for $\pi$, $K$ and $\eta$ are taken as $f_{\pi}=132$ MeV, $f_{K}=f_{\eta}=160$ MeV, respectively. $E_i$ and $M_{i}$ ($E_f$ and $M_f$) are the energy and
mass of the initial (final) heavy hadron, respectively. $\mu_q$ is defined by $1/\mu_q=1/m_j+1/m'_j$, where
$m_j$ and $m'_j$ are the masses of the $j$-th light quark in the initial and final heavy hadrons, respectively.

It is noteworthy that $\mathbf{{\cal H}}_{I}^{RC}$ has often been overlooked in the existing literature~\cite{Li:1994cy,Li:1997gd,Zhao:2002id,DiPierro:2001dwf,Zhong:2008kd,Zhong:2007gp,Ni:2021pce,Xiao:2014ura,Zhong:2010vq,Zhong:2009sk,li:2021hss}.
However, recent investigations into the strong decays of baryons~\cite{Arifi:2021orx,Arifi:2022ntc} have underscored the significance of $\mathbf{{\cal H}}_{I}^{RC}$.
This work signifies the inaugural inclusion of this term within the meson sector, and we will illustrate its pivotal role in accurately determining the widths of mesons.
This effectively addresses the {\bf P2} concern, especially the correct width,  raised on the first page.
	
As we know the vertices described by $\mathbf{{\cal H}}_{I}$
are only effective in the non-perturbative region, which reflect the ability of
$q\bar{q}$ creation in the vacuum. This ability will be suppressed in the
high momentum region due to the weak interactions between the valence quarks.
To suppress the nonphysical contributions in the high momentum region
due to the hard vertices given by the chiral quark model, as indicated in {\bf P3},
we incorporate a suppressed factor $e^{-q^2/(2 \Lambda^2)}$ into the transition amplitude,
\begin{equation}\label{Form factor}
\langle BC,\boldsymbol{q}|{\cal H}_{I} |A\rangle
		\to
\langle BC,\boldsymbol{q}|{\cal H}_{I}e^{-\frac{q^2}{2 \Lambda^2}}| A\rangle,
\end{equation}
as that done in the literature~\cite{Silvestre-Brac:1991qqx,Ortega:2016pgg,Ortega:2016mms,Yang:2021tvc,Yang:2022vdb,Zhong:2022cjx}.
In this study, we employ a fixed cut-off parameter $\Lambda=0.78$ GeV, which is comparable to the value of $0.84$ GeV used in Refs.~\cite{Ortega:2016mms,Ortega:2016pgg}. We should emphasize that the application of suppressed factor is 
another crucial step for obtaining a successful description of
the correct masses for the heavy-light mesons within a unified unquenched quark model as the {\bf P2} concern. 
With this method, the mass corrections contributed by the higher partial wave couplings can be controlled.

Within this framework, we systematically address the significant issues {\bf P1-P3} outlined on the first Page.
	
\begin{figure}
\centering \epsfxsize=8.0 cm \epsfbox{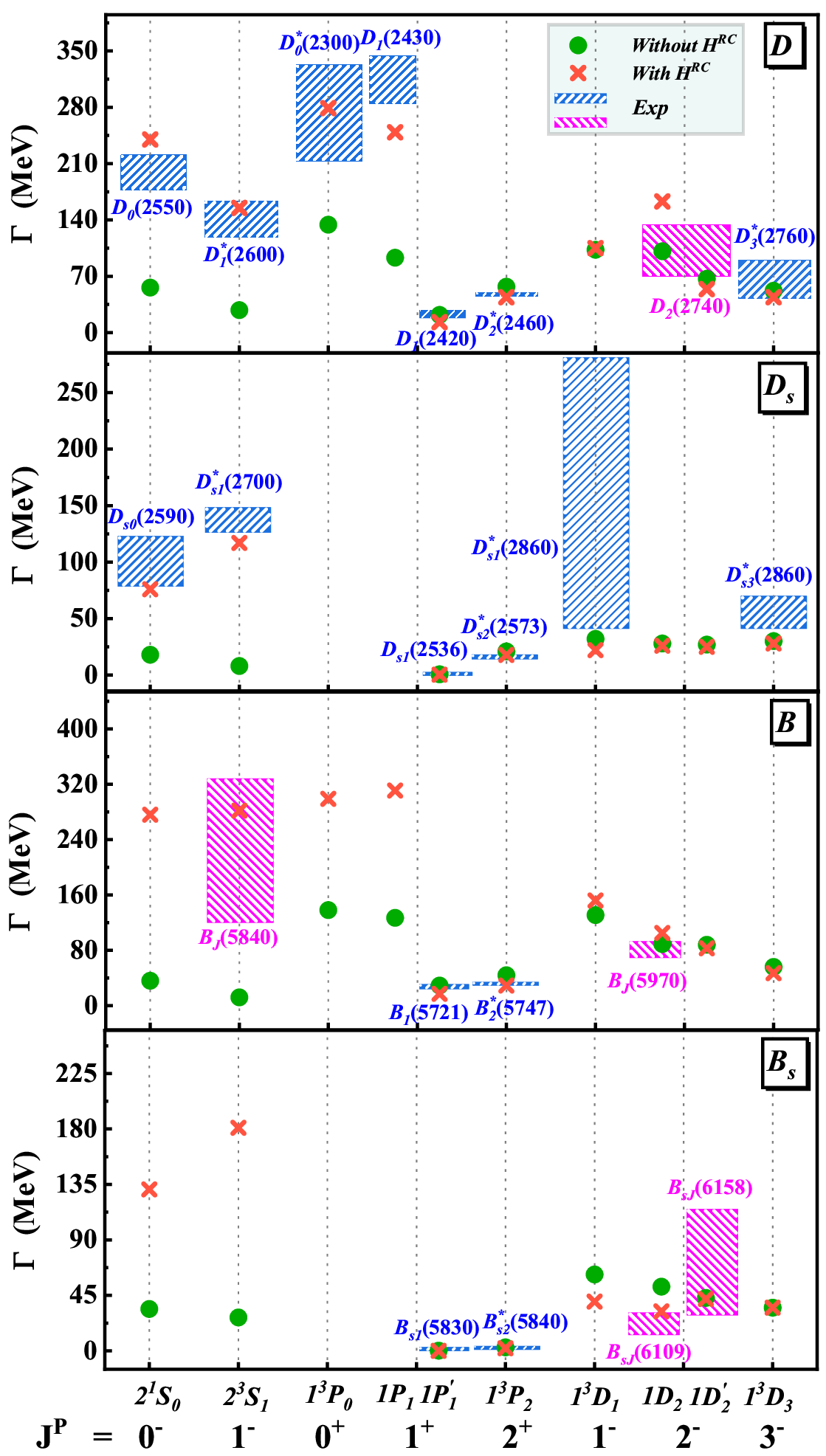} \vspace{-0.2 cm} \caption{Predictions of strong decay widths for the heavy-light mesons compared with the observations~\cite{ParticleDataGroup:2022pth}. The cross (circular) symbol represents the predictions of the width with (without) the incorporation of the relativistic term $\mathcal{H}_{I}^{RC}$ by combining unquenched spectra. }\label{HLwidth}
\end{figure}
	
\begin{figure*}
\centering \epsfxsize=16.0 cm \epsfbox{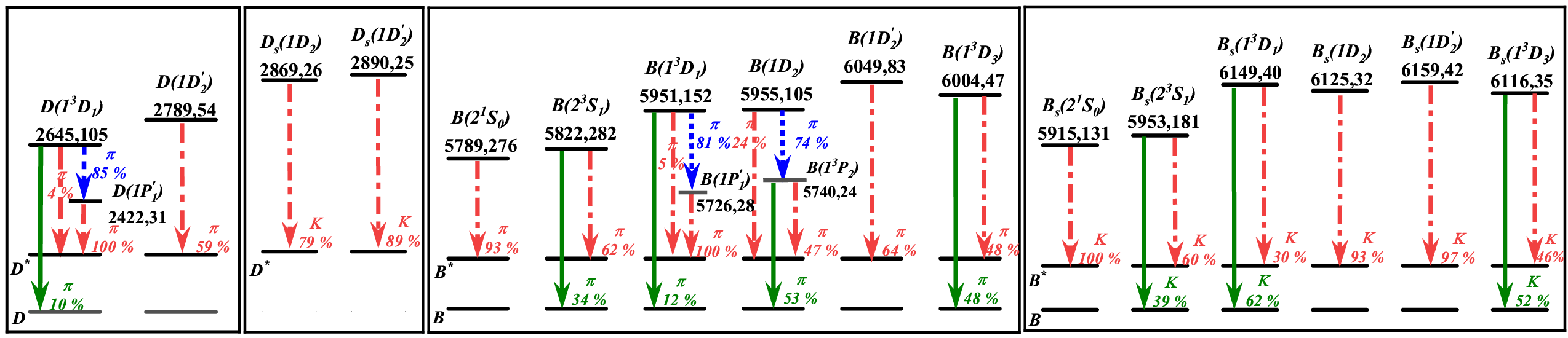} \vspace{-0.2 cm} \caption{Predictions of decay rates of the main decay channels for the missing states in the $D$, $D_s$, $B$, and $B_s$ families.
The values located below the meson state represent the predicted mass and strong decay width in MeV units.}\label{HLR}
\end{figure*}
	
\section{Spectrum}
In Fig.~\ref{Mspectrum}, we present a comprehensive heavy-light meson spectrum across three distinct models alongside existing experimental data.
The crossing points showing the spectrum including the coupled-channel effects are much closer to the experimental data than the results of other models.
This improvement is further illustrated by the significant reduction in the magnitude of  $\chi^2$, defined as $\sum_{i}(\mathrm{Thy}(i)-\mathrm{Exp}(i))^2
	/\mathrm{Error}(i)^2$.
Based on these results, there is no doubt that the coupled channels play a key role in interpolating the spectra of the heavy-light mesons.

Firstly, the coupled-channel effects show their most significant influence on the $^3P_0$ and $P_1$ states.
This is due to their coupling to $S$-wave interactions, involving a pseudo-scalar-heavy-meson and a vector-heavy-meson with a pseudo-scalar-light-meson for $^3P_0$ and $P_1$ states, respectively.
For instance, there are strong interactions between $D_{s0}^*(2317)$ and $DK$, as well as between $D_{s1}(2460)$ and $D^*K$.
These substantial coupled-channel effects naturally account for the significant mass shifts observed in these positive parity states.
In contrast, the other two positive parity states, $P_1'$ and $^3P_2$, which involve $D$-wave interactions with the coupled channels, have minor mass shifts.
This statement is consistent with recent work in Ref.~\cite{Yang:2021tvc}.
The mass of $D^*_0(2300)$ with $J^P=0^+$, which is approximately $2230$ MeV in our results, is slightly lower than the experimental value of $2343\pm 10$ MeV~\cite{ParticleDataGroup:2022pth}.
However, the mass of $D_{0}^*(2300)$ remains a subject of debate in various theoretical studies
~\cite{Du:2020pui,Du:2017zvv,Albaladejo:2016lbb,Bardeen:2003kt,vanBeveren:2003kd}, primarily because its lineshape cannot be explained by the Breit-Wigner form used in the experimental analysis for extracting its mass.
Notably, Lattice QCD calculations are consistent with our findings, where the complex pole position of $D_0^*$ state is at $M-i \Gamma/2=2200 -i 200$ MeV~\cite{Gayer:2021xzv}.

Secondly, the ground states of pseudo-scalars and vectors with $J^P=0^-, 1^-$ exhibit minimal changes across all model for small couple channel effect, because their low mass forbid the decay channels except $D^*$ which also has a very small width.
Conversely, the radial excited states have substantial mass shifts.
The vector-heavy-meson and pseudo-scalar channels will coupled with these excited states in $P$-wave, while for the $D$ and $B$ case, additional positive parity heavy-meson and pseudo-scalar channels can also coupled with them in $S$-wave.
Consequently, the mass shifts are even larger in the $D$ and $B$ cases.
		
Thirdly, let's consider $1D$-wave heavy-light meson states.
It is interesting to note that in the $D_s$ and $B_s$ sectors, the mass shifts are negligible, while they are significant in the $D$ and $B$ sectors.
Actually, for the excited $D_s$ and $B_s$ states, pion production is prohibited due to isospin conservation in strong decay, while $K$ meson is too heavy to effectively couple with positive parity excited $D$ or $B$ states in $S$-wave.
As a result, for $D_s$ and $B_s$ sectors, the coupled-channel effects predominantly involve interactions with higher partial waves.
In contrast,  $S$-wave channels remain available for $D$ and $B$ excited states when coupled with positive parity excited $D$ and $B$ states and pion.
It is natural to interpolate the large mass shift in $D$ and $B$ sectors for $1D$-wave states.

\section{Widths}	
With the parameter set for the mass spectrum, we can compute the strong decay widths for all heavy-light meson resonances.
In Fig.~\ref{HLwidth}, we present the width values with and without $\mathbf{{\cal H}}_{I}^{RC}$, represented by the red crossing and green circle points, respectively, as well as the experimental data.
The widths of all well-established heavy-light meson states can be globally described well by including the relativistic correction term $\mathbf{{\cal H}}_{I}^{RC}$.

For the radially excited states $2^1S_0$ and $2^3S_1$ across all heavy-light mesons, the widths enhance significantly by including $\mathbf{{\cal H}}_{I}^{RC}$.
This naturally addresses a long-standing puzzle, a broad-width of $D_0(2550)$/$D_{s0}(2590)$ and $D^*_1(2600)$/$D^*_{s1}(2700)$ in chiral quark model studies~\cite{Zhong:2008kd,Ni:2021pce,Xiao:2014ura,Zhong:2010vq,Zhong:2009sk}.
For $D^*_{1}(2600)$, the predicted partial width ratio
$R_{th}^{D\pi/D^*\pi}=\Gamma(D\pi)/\Gamma(D^*\pi)\simeq 0.34$ aligns well with the data $R_{exp}^{D\pi/D^*\pi}=0.32\pm 0.11$~\cite{ParticleDataGroup:2022pth}.
Additionally, for the $^3P_0$ and $P'_1$ in $D$ and $B$ sectors, the widths also exhibit increasing, which leads to the predicted widths of $D_0^*(2300)$ and $D_1(2430)$ are both closer to the observations.
These enhancements can be attributed to the second term of $\mathbf{{\cal H}}_{I}^{RC}$, where even powers of quark momentum strengthen the transition, while all other odd-power momentum operators cancel each other~\cite{Arifi:2021orx,Arifi:2022ntc}.
	
\section{Prediction}	
We have demonstrated that our innovative model successfully describes the masses and widths of existing heavy-light mesons.
To further examine our model, we also provide predictions for various masses and widths for heavy-light mesons.	
Upon the confirmation of these states, our understanding of the structure of hadrons will be significantly enriched.
Additionally, we provide the main decay ratios of these states in Fig.~\ref{HLR}.

For the charmed meson, from Fig.~\ref{Mspectrum}, two $^3D_1$ and $D_2'$ states for $D$ meson and two $D_2$
states in $D_s$ sector are all unobserved in the experiments.
These states exhibit narrow widths as illustrated in Fig.~\ref{HLwidth}.
In our model, we also provide the decay rates for these predicted states, as depicted in Fig.~\ref{HLR}.
Notably, for $^3D_1$ $D$ state, the primary decay channel with the widest width is $D_1(2420)\pi$.
This decay mode offers a valuable final state for experimental searches targeting this state.

On the other hand, several excited states of $B$ and $B_s$ mesons remain unobserved, with notable examples being the $^3P_0$ and $P_1'$ states of $B$ and $B_s$.
In our model, we predict the masses of $B(1^3P_0)$ and $B(1P_1)$ to be $5616$ MeV and $5691$ MeV, respectively, which is consistent with those in Refs.~\cite{Bardeen:2003kt,WooLee:2006kdh,Vijande:2007ke}, while $B_s(1^3P_0)$ and $B_s(1P_1)$ are predicted to
be $5711$ MeV and $5781$ MeV here, which are in good agreement with those from Lattice QCD~\cite{Lang:2015hza,Gregory:2010gm}
and other coupled-channel models~\cite{Cheng:2014bca,Cheng:2017oqh,Guo:2006rp,Guo:2006fu, Ortega:2016pgg,Yang:2022vdb,Albaladejo:2016ztm}.
It strengthens the reliability of our model, and the widths of them are also predicted.
As shown in Fig.~\ref{HLR}, the main decay channels of $B(1^3P_0)$ and $B(1P_1)$ are $B\pi$ and $B^*\pi$, respectively, while $B_s(1^3P_0)$ and $B_s(1P_1)$ should be both narrow states which is similar as $D^*_{s0}(2317)$ and $D_{s1}(2460)$.
		
Additionally, numerous negative parity beauty mesons remain to be discovered in experimental searches.
We would like to highlight a few notable cases.
By comparing with the masses, we find that the $B_J(5840)$ favors the $2^3S_1$ state, which is consistent with the assignment in Ref.~\cite{Yu:2019iwm}.
The partial width ratio $\Gamma(B\pi)/\Gamma(B^*\pi)\simeq 0.55$ can be subject to verification in future experiments.
The newly observed structures, $B_{sJ}(6064)$ and $B_{sJ}(6114)$,
in the $B^{+} K^{-}$ mass spectrum at LHCb~\cite{LHCb:2020pet}
, may potentially be explained by the presence of $B_s(1^3D_3)$ and $B_s(1^3D_1)$ states.
Their primary predicted decay channel is $B^{+} K^{-}$.
In addition, these two structures could also be caused by higher mass resonances $B_{sJ}(6109)$ and $B_{sJ}(6158)$, respectively, which mainly decay into $B^{*+} K^{-}$.

\section{Summary}		
	
In this study, we propose a unified unquenched quark model framework, which for the first time successfully provides 
comprehensive mass spectra, decay widths, and strong decay branch widths for the $D$, $D_s$, $B$, and $B_s$ sectors.
This unquenched quark model not only combines the traditional quark model and the coupled-channel effects based on chiral dynamics,
but also incorporates a relativistic correction term for describing the strong transition 
amplitude for the first time. These theoretical approaches and strategies adopted in the present work ensure that our model is 
excellent for simultaneously describing the masses and decay widths of all existing heavy-light mesons.

Notably, our model naturally resolves the long-standing puzzles that the low-mass nature of $D^*_{s0}(2317)$ and $D_{s1}(2460)$ and the broad nature of radial excitations $D_0(2550)$/$D_{s0}(2590)$ and $D^*_1(2600)$/$D^*_{s1}(2700)$.
Furthermore, our model predicts the masses, widths, and branching ratios of various new states, offering valuable guidance for their discovery in future experiments. Some of our predictions are consistent with the lattice data, strengthening our confidence in extending this model to the other hadron spectra, such as the light baryon spectrum and singly-heavy baryon spectrum. 
The success of the unquenched quark model presented in this work indicates 
it may be an important step for understanding the hadron spectrum.

\section*{Acknowledgements }
	
We thank useful discussions from Qiang Zhao, Xiang Liu, Zhi-Yong Zhou, and A. J. Arifi.
This work is supported by the National Natural Science Foundation of China under Grants Nos.12175065 and 12235018 (X.H.Z), 12175239
and 12221005 (J.J.W), and by the National Key R $\&$ D Program of China
under Contract No. 2020YFA0406400 (J.J.W),
and supported by Chinese Academy of Sciences under Grant No. YSBR-101 (J.J.W).
	


\bibliographystyle{unsrt}

\appendix

\section*{Supplemental Material}

\subsection{Potential model}

The bare mass and numerical wave functions of the heavy-light meson are calculate by a semirelativistic potential quark model.
In the model, the effective Hamiltonian is given by
\begin{equation}\label{hpoten1}
 \mathcal{H}_0=\sqrt{\mathbf{p}_1^2+m_1^2}+\sqrt{\mathbf{p}_2^2+m_2^2}+V(r),
\end{equation}
where the first two terms stand for the kinetic energies for the light antiquark
$\bar{q}$ and heavy quark $Q$, respectively, $V(r)$ stands for the effective potentials
between two quarks, which is given by
\begin{equation}\label{hpoten}
 V(r)=V_0(r)+V_{sd}(r).
\end{equation}
Here $V_0(r)$ is the well-known Cornell potential~\cite{Eichten:1978tg}
\begin{eqnarray}\label{H0}
V_0(r)=-\frac{4}{3}\frac{\alpha_s(r)}{r}+br+C_{0},
\end{eqnarray}
which includes the color Coulomb interaction and linear confinement, and zero point energy $C_0$. The $V_{sd}(r)$ is the spin-dependent part, we adopted the widely used form~\cite{Godfrey:1985xj,Swanson:2005}
\begin{eqnarray}\label{Hsd}
\begin{aligned}
V_{sd}(r)&=\frac{32\pi \alpha_{s}(r) \cdot \sigma^3 e^{-\sigma^2 r^2}}{9 \sqrt{\pi}  \tilde{m}_1m_2} \mathbf{S_1} \cdot \mathbf{S_2} \\
 &+\frac{4}{3}\frac{\alpha_s(r)}{\tilde{m}_1 m_2}\frac{1}{r^3}\left(\frac{3\mathbf{S}_{1}\cdot \mathbf{r}\mathbf{S}_{2}\cdot \mathbf{r}}{r^2}-\mathbf{S}_{1}\cdot\mathbf{S}_{2}\right)+H_{LS}.
\end{aligned}
\end{eqnarray}
In the Eq.~(\ref{Hsd}), the first term is the spin-spin contact hyperfine potential, the second term is the tensor potential,
the $H_{LS}$ stands for the spin-orbit interaction, which can be decomposed into
symmetric part $H_{sym}$ and antisymmetric part $H_{anti}$:
\begin{eqnarray}\label{vs}
H_{sym}=\frac{\mathbf{S_{+}\cdot L}}{2}\left[ \left(\frac{1}{2  \tilde{m}_1^2}+\frac{1}{2m_2^2} \right) \left( \frac{4 \alpha_s(r)}{3r^3}-\frac{b}{r}   \right)+\frac{8 \alpha_s(r)}{3  \tilde{m}_1 m_2r^3} \right],\\
H_{anti}=\frac{\mathbf{S_{-}\cdot L}}{2}\left(\frac{1}{2  \tilde{m}_1^2}-\frac{1}{2m_2^2} \right) \left( \frac{4 \alpha_s(r)}{3r^3}-\frac{b}{r}   \right) .\ \ \ \ \ \ \ \ \ \ \ \ \ \ \ \ \ \ \ \ \ \ \
\end{eqnarray}
In these equations, $\mathbf{L}$ is the relative orbital angular momentum of the $Q\bar{q}$
system; $\mathbf{S}_1$ and $\mathbf{S}_{2}$ are the spins of the light and heavy quarks, respectively, and $\mathbf{S}_{\pm}\equiv\mathbf{S}_1\pm \mathbf{S}_{2}$.

The antisymmetric part of the spin-orbit interaction, $H_{anti}$, can cause a
configuration mixing between spin-triplet $n^{3}L_{J}$ and spin-singlet $n^{1}L_{J}$
for the $Q\bar{q}$ system. Thus, the physical states $nL_J$ and $nL'_J$ are expressed as
\begin{equation}\label{mixsta}
\left(
  \begin{array}{c}
   nL_J\\
   nL'_J\\
  \end{array}\right)=
  \left(
  \begin{array}{cc}
   \cos\theta_{nL} &\sin\theta_{nL}\\
  -\sin\theta_{nL} &\cos\theta_{nL}\\
  \end{array}
\right)
\left(
  \begin{array}{c}
  n^{1}L_{J}\\
  n^{3}L_{J}\\
  \end{array}\right),
\end{equation}
where $J=L=1,2,3\cdots$, and the $\theta_{nL}$ is the mixing angle.
In this work, $nL_J$ and $nL'_J$ correspond to the lower-mass and higher-mass mixed states, respectively, as often adopted in the literature.
This mixing angle can be perturbatively determined with the nondiagonal matrix element $\langle n^{1}L_{J} |\mathcal{H}_{anti}| n^{3}L_{J} \rangle$.

\begin{table*}
\begin{center}
\caption{Potential model parameters for quenched and unquenched mass spectra.}\label{Parameters}
\begin{tabular}{cccccccccccccccc}
\bottomrule[1.0pt]\bottomrule[1.0pt]
&$m_{c/b}$(GeV)   &$m_{u,d}$(GeV)   &$\tilde{m}_{u,d}$(GeV)   &$m_s$(GeV)   &$\tilde{m}_s$(GeV)   &$\alpha_1$   &$b$$(\textrm{GeV}^2)$      &$\sigma$(GeV)   &$C_0$(MeV)   &$r_c$(fm) \\
\bottomrule[0.5pt]
\multicolumn{11}{l}{\emph{quenched}}        \\
$D$       &$1.45$    &$0.45$        &$0.64$        &$...$        &$...$       &$0.28$    &$0.180$      &$0.953$      &$-302.0$    &$0.332$  \\
$D_s$     &$1.45$    &$...$         &$...$        &$0.55$        &$0.70$      &$0.28$    &$0.180$      &$0.990$      &$-267.0$    &$0.320$ \\
$B$       &$4.80$    &$0.45$        &$0.64$        &$...$        &$...$       &$0.22$    &$0.180$      &$0.870$      &$-246.0$    &$0.266$  \\
$B_s$     &$4.80$    &$...$         &$...$        &$0.55$        &$0.70$      &$0.22$    &$0.180$      &$0.965$      &$-218.0$    &$0.253$ \\
\bottomrule[0.5pt]
\multicolumn{11}{l}{\emph{unquenched}}        \\
$D$       &$1.45$    &$0.35$        &$0.64$        &$...$        &$...$       &$0.38$    &$0.180$      &$0.930$      &$-179.0$    &$0.341$  \\
$D_s$     &$1.45$    &$...$         &$...$         &$0.55$       &$0.70$      &$0.38$    &$0.180$      &$0.882$      &$-212.0$    &$0.325$ \\
$B$       &$4.80$    &$0.35$        &$0.64$        &$...$        &$...$       &$0.25$    &$0.180$      &$0.870$      &$-172.0$    &$0.270$  \\
$B_s$     &$4.80$    &$...$         &$...$         &$0.55$       &$0.70$      &$0.25$    &$0.180$      &$0.939$      &$-200.0$    &$0.256$ \\
\bottomrule[1.0pt]\bottomrule[1.0pt]
\end{tabular}
\end{center}
\end{table*}

The running coupling constant $\alpha_s(r)$ adopted a parameterized form, $\alpha_s(r) =\sum_{i=1,2,3} \alpha_i \frac{2}{\sqrt{\pi}}\int_0^{\gamma_i r}e^{-x^2}dx$, in the coordinate space.
In the different heavy-light meson systems, we set the consistent parameters $\alpha_2=0.15$, $\alpha_3=0.20$, $\gamma_1=1/2$, $\gamma_2=\sqrt{10}/2$, and $\gamma_3=\sqrt{1000}/2$, which are the same as those adopted in Refs.~\cite{Godfrey:1985xj,Liu:2013maa,Liu:2015lka}.
The parameter $\alpha_1$ is slightly different for $D_{(s)}$ and $B_{(s)}$ spectra
for a better description the mass spectrum, as shown in the value in Table~\ref{Parameters}.
It should be mentioned that in the spin-dependent potentials, we have
replaced the light quark mass $m_1$ with $\tilde{m}_1$ to include some
relativistic corrections.
By using the Gaussian expansion method~\cite{Hiyama:2003cu} to solve the Schr$\mathrm{\ddot{o}}$dinger equation, the bare mass and numerical wave functions for heavy-light mesons are derived.
Detailed discussions on the numerical calculation techniques involving Gaussian expansion can be found in our previous work~\cite{Ni:2021pce}.

\begin{table}
\begin{center}
\caption{The effective $\beta_{eff}$~(GeV) parameters for the ground states $D^{(*)}$ and $B^{(*)}$ in the quenched/unquenched pictures. In the table,
$\beta_{eff}^A$ stands for the results determined by the strong decay properties of the well-established heavy-light states, while
$\beta_{eff}^B$ stands for the results extracted from the numerical wave functions the ground states $D^{(*)}$ and $B^{(*)}$ obtained in the potential model.  }\label{effbeta}
\scalebox{1.0}{
\begin{tabular}{ccccccccc}
\bottomrule[1.0pt]\bottomrule[1.0pt]
    $$    ~~~    &$D$      ~~~   &$D^*$    ~~~  &$B$      ~~~   &$B^*$    \\
  \bottomrule[0.8pt]
  $\beta_{eff}^A$
  &$0.499/0.497$  ~~~&$0.454/0.449$  ~~~&$0.524/0.509$  ~~~&$0.512/0.498$    \\
  $\beta_{eff}^B$
  &$0.587/0.585$  ~~~&$0.493/0.488$  ~~~&$0.616/0.599$  ~~~&$0.582/0.566$  \\
\bottomrule[1.0pt]\bottomrule[1.0pt]
\end{tabular}}
\end{center}
\end{table}

\subsection{Model parameters}

In the quark potential model, the slope parameter $b$ for the linear potential is taken to be $b=0.18~\mathrm{GeV}^2$, which is a typical value adopted in various relativistic potential models~\cite{Godfrey:1985xj,Zeng:1994vj,Ebert:1997nk}.
The other parameters, $\{\alpha_{1}, \sigma, C_{0}, m_{c}, m_{b}, m_{u/d}, m_{s}, \tilde{m}_{u/d}, \tilde{m}_{s}, r_c \}$, are determined by fitting the low-lying well-established states. To overcome the singular behavior of $1/r^3$ in the spin-dependent potentials,
we introduce a cutoff distance $r_c$ in the
calculation, which allows $1/r^3=1/r_c^3$ within a small range $r\in (0,r_c)$, as we did in the
previous studies~\cite{Deng:2016ktl,Deng:2016stx,li:2021hss,Ni:2021pce}.
In this work, we give two solutions for the heavy-light meson spectrum.
One solution is obtained within the quenched quark model without coupled-channel effects, and the other one is obtained within the unquenched quark model with coupled-channel effects.
The parameter sets have been listed in Table~\ref{Parameters}.
For the calculation of the strong transition amplitude $\langle BC,\boldsymbol{q}|{\mathcal{H}}_{I}|A\rangle$ within the chiral quark model, one can find that the relativistic correction term $\mathcal{H}_{I}^{RC}$ is sensitive to the effective mass of the light $u$/$d$ quarks due to the factor
$1/\mu_q^2 = \left( 1/m_j + 1/m'_j \right)^2$.
Thus, to obtain a good description of the strong decay properties and coupled-channel effects for the heavy-light meson states, the effective mass
of the $u$/$d$ quark is taken to be $m_u = m_d = 0.45$ GeV, which is slightly larger than that used in the potential model for the study of the mass spectrum.
In fact, the operator $\mathcal{H}_{I}$ is derived from the effective Lagrangians through a weak binding approximation, the interaction between the quarks is not seriously considered as that in the potential model.
Thus, the effective $u$/$d$ quark mass in the chiral quark model should be slightly different from that in the potential model.

Furthermore, to obtain the transition amplitude $\langle BC,\boldsymbol{q}|{\mathcal{H}}_{I}|A\rangle$, we adopt the numerical wave functions for the heavy-light mesons, except the ground states $D^{(*)}$ and $B^{(*)}$.
We have noted that the decays of most excited heavy-light states are related to the ground states, their transition amplitudes are sensitive to the details of the wave functions of the ground states $D^{(*)}$ and $B^{(*)}$.
Considering the uncertainty of the ground state wave functions, we properly adjust their size to more reasonably describe the strong decay properties of the well-established states, such as $D_{2}^{*}(2460)$, $D_{s2}^{*}(2573)$, $B_{2}^{*}(5747)$, and $B_{s2}^{*}(5840)$.
For convenience, the ground wave functions are adopted a simple harmonic oscillator form with an effective size parameter $\beta_{eff}$.
The determined $\beta_{eff}$ parameters for $D^{(*)}$ and $B^{(*)}$ have been listed in Table~\ref{effbeta}.
There is about a $10 \%$ correction to the effective $\beta_{eff}$ parameters which are extracted from the numerical wave functions of $D^{(*)}$ and $B^{(*)}$ by reproducing the root-mean-square radius $\sqrt{\langle r^2 \rangle}$ with a simple harmonic oscillator form.

\subsection{More details of numerical results}
The masses for the heavy-light meson states obtained from the quenched and unquenched quark models are listed in Table~\ref{Mass and Width}.
The decay widths of the heavy-light meson from the two methods are also listed in Table~\ref{Mass and Width}.
In method I, the decay widths are described without relativistic correction term $\mathcal{H}_{I}^{RC}$ by combining the spectrum from the unquenched quark model.
While in method II, the decay widths are described with relativistic correction term $\mathcal{H}_{I}^{RC}$ by combining
the spectrum also from the unquenched quark model.
It is found that with relativistic correction of the descriptions of the decay widths for the heavy-light states have an overall improvement.

In Tables~\ref{charmdecay} and \ref{BBsmeson}, the mass shifts, and partial widths of the heavy light meson states contributed by each channel are given.
From the tables, one can find which channels play a crucial role in the mass shift of a heavy-light meson, and which channels dominate their strong decays.
One also can see the details of the relativistic corrections to the partial widths of each channel.
The partial width ratios between different decay channels and their branching fractions for each state can be obtained from these tables.

\begin{table*}
\begin{center}
\renewcommand\arraystretch{1.15}
\tabcolsep=0.060cm
\caption{Theoretical masses and widths compared with the data. The mass spectra from both the quenched and unquenched pictures are given, which are denoted with $Q$ and $UQ$, respectively.
The strong decay widths, which combine the unquenched spectra and are described by non-relativistic chiral interactions $\mathcal{H}_{I}^{NR}$, are further augmented by a relativistic correction term $\mathcal{H}_{I}^{RC}$. These widths are denoted as $\Gamma_i^{NR}$ and $\Gamma_i^{NR+RC}$, respectively. The experimental data are taken from the PDG~\cite{ParticleDataGroup:2022pth}. The mixing angles
for $1P_1$/$1P_1'$ states in the $D$, $D_s$, $B$ and $B_s$ families are determined to be $-30.0^\circ$,
$-26.5^\circ$, $-34.3^\circ$, and $-34.2^\circ$, respectively. While the mixing angles
for $1D_2$/$1D_2'$ states in the $D$, $D_s$, $B$ and $B_s$ families are determined to be $-40.5^\circ$,
$-40.5^\circ$, $-39.9^\circ$, and $-40.2^\circ$, respectively. }\label{Mass and Width}
\begin{tabular*}{179mm}{cccccccccccccccccccccccccccccccc}
\bottomrule[1.0pt]\bottomrule[1.0pt]
&\multicolumn{8}{c}{$D~mesons$}
&
&\multicolumn{8}{c}{$D_s~mesons$}  \\
\cline{2-9} \cline{11-18}
& Observed
&\multicolumn{3}{c}{Mass~(MeV)}
&
&\multicolumn{3}{c}{Width~(MeV)}
&
&Observed
&\multicolumn{3}{c}{Mass~(MeV)}
&
&\multicolumn{3}{c}{Width~(MeV)}
\\
\cline{3-5} \cline{7-9} \cline{12-14} \cline{16-18}
$n^{2S+1}L_J$
&state
&\emph{Q}
&\emph{UQ}
&\emph{Exp.}
&
&\emph{$\Gamma_i^{NR}$}
&\emph{$\Gamma_i^{NR+RC}$}
&\emph{Exp.}
&
&state
&\emph{Q}
&\emph{UQ}
&\emph{Exp.}
&
&\emph{$\Gamma_i^{NR}$}
&\emph{$\Gamma_i^{NR+RC}$}
&\emph{Exp.}
\\
\bottomrule[0.7pt]
$1^1S_0$          &$D^0$                    &$1865$   &$1865$      &$1865$                 & &$-$    &$-$       &$-$                 & &$D_s$            &$1969$   &$1969$     &$1969$          &  &$-$   &$-$  &$-$         \\
$1^3S_1$          &$D^*$                    &$2010$   &$2010$      &$2008$                 & &$0.0647$    &$0.0183$       &$<2.1$                 & &$D_s^*$            &$2112$   &$2112$     &$2112$          &  &$-$   &$-$  &$-$         \\
&$    $                   &$    $   &$    $      &$2010$                 & &$0.1321$   &$0.0375$       &$0.0834$                & &$    $             &$    $   &$    $     &$    $          &  &$ $   &$ $  &$ $         \\
\bottomrule[0.7pt]
$2^1S_0$          &$D_0(2550)$              &$2554$   &$2502$      &$2518\pm 9$            & &$56$      &$240$        &$199\pm 22$               & &$D_{s0}(2590)$     &$2655$   &$2629$     &$2591\pm 9$     &  &$18$   &$76$   &$89\pm  20$    \\
$2^3S_1$          &$D_1^*(2600)$            &$2640$   &$2602$      &$2627\pm 10$           & &$28$      &$155$        &$141\pm 23$               & &$D_{s1}^*(2700)$   &$2738$   &$2743$     &$2714\pm 5$     &  &$8$    &$117$  &$122\pm 10$    \\
\cline{1-9} \cline{11-18}
$1^3P_0$          &$D_0^*(2300)$            &$2307$   &$2226$      &$2297\pm 28$           & &$134$     &$279$        &$273\pm 60$               & &$D_{s0}^*(2317)$   &$2400$   &$2312$     &$2318$          &  &$-$    &$-$  &$<3.8$          \\
$1P_1$            &$D_1(2430)$              &$2422$   &$2410$      &$2412\pm 9$            & &$93$      &$249$        &$314\pm 29$               & &$D_{s1}(2460)$     &$2522$   &$2484$     &$2460$          &  &$-$    &$-$  &$<3.5$          \\
$1P'_1$           &$D_1(2420)$              &$2453$   &$2435$      &$2427.2\pm 2.2$        & &$21.5$    &$12.5$       &$23.2\pm 4.6$             & &$D_{s1}(2536)$     &$2544$   &$2518$     &$2535$          &  &$0.6$  &$0.4$  &$0.92\pm0.05$   \\
$1^3P_2$          &$D_2^*(2460)$            &$2475$   &$2464$      &$2461.1\pm 0.7$        & &$56.7$    &$43.6$       &$47.3\pm 0.8$             & &$D_{s2}^*(2573)$   &$2574$   &$2557$     &$2569$          &  &$21.3$ &$17.9$ &$16.9\pm 0.7$    \\
\bottomrule[0.7pt]
$1^3D_1$          &$-$                      &$2750$   &$2645$      &$-$                    & &$103$     &$105$        &$-$                       & &$D_{s1}^*(2860)$   &$2838$   &$2825$     &$2859\pm41$     &  &$32$   &$22$   &$159\pm 122$     \\
$1D_2$            &$D_2(2740)$              &$2757$   &$2722$      &$2751\pm10$            & &$101$     &$163$        &$102\pm32$                & &$-$                &$2855$   &$2869$     &$-$             &  &$28$   &$26$   &$-$            \\
$1D'_2$           &$-$                      &$2822$   &$2789$      &$-$                    & &$67$     &$54$         &$-$                        & &$-$                &$2907$   &$2890$     &$-$             &  &$27$   &$25$   &$-$            \\
$1^3D_3$          &$D_3^*(2750)$            &$2782$   &$2771$      &$2753\pm 10$           & &$52$      &$44$         &$66\pm 24$                & &$D_{s3}^*(2860)$   &$2879$   &$2888$     &$2860.5\pm11.1$ &  &$30$   &$28$   &$53\pm 17$       \\
\bottomrule[0.7pt]\bottomrule[0.7pt]
&\multicolumn{8}{c}{$B~mesons$}
&
&\multicolumn{8}{c}{$B_s~mesons$}  \\
\cline{2-9} \cline{11-18}
&Observed
&\multicolumn{3}{c}{Mass~(MeV)}
&
&\multicolumn{3}{c}{Width~(MeV)}
&
&Observed
&\multicolumn{3}{c}{Mass~(MeV)}
&
&\multicolumn{3}{c}{Width~(MeV)}
\\
\cline{3-5} \cline{7-9} \cline{12-14} \cline{16-18}
$n^{2S+1}L_J$
&state
&\emph{Q}
&\emph{UQ}
&\emph{Exp.}
&
&\emph{$\Gamma_i^{NR}$}
&\emph{$\Gamma_i^{NR+RC}$}
&\emph{Exp.}
&
&state
&\emph{Q}
&\emph{UQ}
&\emph{Exp.}
&
&\emph{$\Gamma_i^{NR}$}
&\emph{$\Gamma_i^{NR+RC}$}
&\emph{Exp.}
\\
\bottomrule[0.7pt]
$1^1S_0$          &$B^0$                    &$5280$   &$5280$      &$5280$                 & &$-$    &$-$       &$-$                 & &$B_s$            &$5367$   &$5367$     &$5367$          &  &$-$   &$-$  &$-$         \\
$1^3S_1$ &$B^*$           &$5325$  &$5325$      &$5325$                   & &$-$       &$-$    &$-$                                 &  &$B_s^*$            &$5416$ &$5416$      &$5416$            & &$-$     &$-$     &$-$      \\
$2^1S_0$ &$-$             &$5877$  &$5789$      &$-$                      & &$36$      &$276$  &$-$                                 &  &$-$                &$5964$ &$5915$      &$-$               & &$34$    &$131$   &$-$      \\
$2^3S_1$ &$B_J(5840)^+$?  &$5902$  &$5822$      &$5851\pm 19$             & &$12$       &$282$  &${224\pm 104}$                     &  &$-$                &$5991$ &$5953$      &$-$               & &$27$    &$181$   &$-$      \\
\bottomrule[0.7pt]
$1^3P_0$ &$-$             &$5695$  &$5616$      &$-$                      & &$138$     &$299$  &$-$                                 &  &$-$                &$5775$ &$5711$      &$-$               & &$-$     &$-$     &$-$       \\
$1P_1$   &$-$             &$5746$  &$5691$      &$-$                      & &$127$     &$311$  &$-$                                 &  &$-$                &$5834$ &$5781$      &$-$               & &$-$     &$-$     &$-$       \\
$1P'_1$  &$B_1(5721)^0$   &$5762$  &$5730$      &$5726.1\pm1.3$           & &$29.0$    &$16.5$ &$27.5\pm 3.4$                       &  &$B_{s1}(5830)$     &$5842$ &$5810$      &$5829$            & &$0.06$  &$0.04$  &$0.5\pm0.4$   \\
$1^3P_2$ &$B_2^*(5747)^0$ &$5769$  &$5734$      &$5739.5\pm0.7$           & &$43.5$    &$29.1$ &$24.2\pm 1.7$                       &  &$B_{s2}^*(5840)$   &$5857$ &$5827$      &$5840$            & &$2.79$  &$2.07$  &$1.49\pm 0.27$ \\
\bottomrule[0.7pt]
$1^3D_1$ &$-$             &$6080$  &$5951$      &$-$                      & &$131$     &$152$  &$-$                                 &  &$-$                &$6154$ &$6149$      &$-$               & &$62$   &$40$    &$-$     \\
$1D_2$   &$B_J(5970)^0$?  &$6036$  &$5955$      &$5971\pm 5$              & &$89$      &$105$  &$81\pm 12$                          &  &$B_{sJ}(6109)$     &$6123$ &$6125$      &$6108.8\pm 1.8$   & &$52$   &$32$    &$22\pm 9$       \\
$1D'_2$  &$-$             &$6114$  &$6049$      &$-$                      & &$88$      &$83$   &$-$                                 &  &$B_{sJ}(6158)$     &$6186$ &$6159$      &$6158\pm 9$       & &$43$   &$42$    &$72\pm 43$          \\
$1^3D_3$ &$-$             &$6051$  &$6004$      &$-$                      & &$56$      &$47$   &$-$                                 &  &$-$                &$6137$ &$6116$      &$-$               & &$35$  &$35$    &$-$      \\
\bottomrule[1.0pt]\bottomrule[1.0pt]
\end{tabular*}
\end{center}
\end{table*}

\begin{table*}
\caption{The mass shifts $\Delta M$ and partial widths (in MeV) of the $D$ and $D_s$ states.
The bare masses obtained from the potential model are listed in square brackets. The strong decay widths, which combine the unquenched spectra and are described by non-relativistic chiral interactions $\mathcal{H}_{I}^{NR}$, are further augmented by a relativistic correction term $\mathcal{H}_{I}^{RC}$. These widths are denoted as $\Gamma_i^{NR}$ and $\Gamma_i^{NR+RC}$, respectively. The forbidden decay channel is denoted by ``...". The experimental data are taken from the PDG~\cite{ParticleDataGroup:2022pth}.} \label{charmdecay}
\renewcommand\arraystretch{1.10}
\tabcolsep=0.182cm
\resizebox{\textwidth}{!}{
\begin{tabular*}{190mm}{ccccccccc|lccccccccc}
\bottomrule[1.0pt]\bottomrule[1.0pt]
                       &\multicolumn{3}{c}{$D(2^1S_0)$}&
&\multicolumn{3}{c}{$D(2^3S_1)$}
&&&
&\multicolumn{3}{c}{$D(1^3P_0)$}&
&\multicolumn{3}{c}{$D(1^3P_2)$}\\
\cline{2-4}  \cline{6-8} \cline{12-14}  \cline{16-18}
&$\Delta M$   &\multicolumn{2}{c}{$as~D_0(2550)$}&
&$\Delta M$   &\multicolumn{2}{c}{$as~D_1^*(2600)$}
&&&
&$\Delta M$  &\multicolumn{2}{c}{$as~D_0^*(2300)$}&
&$\Delta M$  &\multicolumn{2}{c}{$as~D_2^*(2460)$}\\
\cline{3-4}  \cline{7-8} \cline{13-14}  \cline{17-18}
Channel                &$[2575]$    &$\Gamma_i^{NR}$   &$\Gamma_i^{NR+RC}$&
&$[2667]$    &$\Gamma_i^{NR}$   &$\Gamma_i^{NR+RC}$
&&&Channel
&$[2304]$    &$\Gamma_i^{NR}$   &$\Gamma_i^{NR+RC}$&
&$[2510]$    &$\Gamma_i^{NR}$   &$\Gamma_i^{NR+RC}$\\
\cline{1-4}  \cline{6-8} \cline{11-14}  \cline{16-18}
$D\pi$                  &$...$          &$...$        &$...$            &  &$4.4 $           &$11.2$            &$31.0 $              &&& $D\pi$                &$-78.1$              &$134.0$             &$279.4$                   &     &$-19.3$           &$35.7$             &$29.0 $            \\
$D^\ast\pi$             &$-36.2$        &$43.1$       &$225.6$          &  &$-2.1$           &$0.4 $            &$92.0 $              &&& $D^\ast\pi$           &$...$                &$...$               &$...$                     &     &$-24.5$           &$20.9$             &$14.5 $            \\
$D\eta$                 &$...$          &$...$        &$...$            &  &$-0.6$           &$0.02$            &$4.2  $              &&& $D\eta$               &$...$                &$...$               &$...$                     &     &$-2.1$            &$0.1 $             &$0.1  $            \\
$D^\ast\eta$            &$...$          &$...$        &$...$            &  &$-2.2$           &$0.7 $            &$3.9  $              &&&  \emph{Total}         &$\mathbf{-78.1}$     &$\mathbf{134.0}$    &$\mathbf{279.4}$          &     &$\mathbf{-45.9}$  &$\mathbf{56.7}$    &$\mathbf{43.6}$     \\
$D_sK$                  &$...$          &$...$        &$...$            &  &$-1.4$           &$1.8 $            &$4.1  $              &&&  \emph{Exp.}          &$\mathbf{-}$         &\multicolumn{2}{c}{$\mathbf{273\pm60}$}        &     &$\mathbf{-}$      &\multicolumn{2}{c}{$\mathbf{47.3\pm0.8}$}   \\
\cline{11-18}
$D_s^\ast{K}$           &$...$          &$...$        &$...$            &  &$-2.8$           &$0.03$            &$1.0 $               &&&  \multicolumn{8}{c}{\quad}                                                                                                                                 \\
\cline{11-18}
$D_0^\ast(2300)\pi$     &$-36.3$        &$13.0$       &$14.3$           &  &$...$            &$...$             &$...$                &&&                      &\multicolumn{3}{c}{$D(1P_1)$}                      &&\multicolumn{3}{c}{$D(1P'_1)$}                           \\
\cline{12-14}  \cline{16-18}
$D_1(2430)\pi$          &$...$          &$...$        &$...$            &  &$-43.6$          &$14.0$            &$18.9 $              &&&                     &$\Delta M$&\multicolumn{2}{c}{$as~D_1(2430)$}      &&$\Delta M$&\multicolumn{2}{c}{$as~D_1(2420)$}     \\
\cline{13-14}  \cline{17-18}
$D_1(2420)\pi$          &$...$          &$...$        &$...$            &  &$-6.7 $          &$0.2$             &$0.02$               &&&  Channel             &$[2449]$           &$\Gamma_i^{NR}$         &$\Gamma_i^{NR+RC}$    &   &$[2471]$        &$\Gamma_i^{NR}$     &$\Gamma_i^{NR+RC}$         \\
\cline{11-14} \cline{16-18}
$D_2^\ast(2460)\pi$     &$...$         &$...$        &$...$             &  &$-10.4$          &$0.02 $           &$0.01$               &&&  $D^\ast\pi$         &$-39.3$            &$92.9$                  &$249.2$               &   &$-36.1$         &$21.5$              &$12.5$            \\
\emph{Total}            &$\mathbf{-72.5}$    &$\mathbf{56.1}$   &$\mathbf{239.9}$       &  &$\mathbf{-65.4}$ &$\mathbf{28.4}$   &$\mathbf{155.1}$     &&&  \emph{Total}        &$\mathbf{-39.3}$   &$\mathbf{92.9}$         &$\mathbf{249.2}$      &   &$\mathbf{-36.1}$&$\mathbf{21.5}$     &$\mathbf{12.5}$    \\
\emph{Exp.}             &$\mathbf{-}$        &\multicolumn{2}{c}{$\mathbf{199\pm22}$}   &  &$\mathbf{-}$     &\multicolumn{2}{c}{$\mathbf{141\pm23}$} &&&  \emph{Exp.}         &$\mathbf{-}$       &\multicolumn{2}{c}{$\mathbf{314\pm29}$}        &   &$\mathbf{-}$    &\multicolumn{2}{c}{$\mathbf{23.2\pm4.6}$}  \\
\bottomrule[0.7pt]
&\multicolumn{3}{c}{$D(1^3D_1)$}&
&\multicolumn{3}{c}{$D(1^3D_3)$}
&&&
&\multicolumn{3}{c}{$D(1D_2)$}&
&\multicolumn{3}{c}{$D(1D'_2)$}\\
\cline{2-4}  \cline{6-8} \cline{12-14}  \cline{16-18}
&$\Delta M$   &\multicolumn{2}{c}{$M=2645$}&
&$\Delta M$   &\multicolumn{2}{c}{$as~D_3^*(2750)$}
&&&
&$\Delta M$  &\multicolumn{2}{c}{$as~D_2(2740)$}&
&$\Delta M$  &\multicolumn{2}{c}{$M=2789$}                             \\
\cline{3-4}  \cline{7-8} \cline{13-14}  \cline{17-18}
Channel                &$[2765]$  &$\Gamma_i^{NR}$             &$\Gamma_i^{NR+RC}$&
&$[2830]$  &$\Gamma_i^{NR}$             &$\Gamma_i^{NR+RC}$
&&&Channel
&$[2799]$  &$\Gamma_i^{NR}$             &$\Gamma_i^{NR+RC}$&
&$[2856]$  &$\Gamma_i^{NR}$             &$\Gamma_i^{NR+RC}$\\
\cline{1-4}  \cline{6-8} \cline{11-14}  \cline{16-18}
$D\pi$                  &$-2.9$            &$18.7$             &$10.1$             &      &$-7.6  $          &$20.0 $             &$21.2$                 &&&       $D^\ast\pi$             &$-3.9$           &$17.3$           &$23.1$             &      &$-20.1 $          &$36.8 $             &$31.7$        \\
$D^\ast\pi$             &$-1.8$            &$7.8$              &$4.5 $             &      &$-11.5 $          &$18.4 $             &$17.4$                 &&&       $D^\ast\eta$            &$-0.8$           &$4.7$            &$2.1$              &      &$-2.5  $          &$1.3  $             &$1.0 $        \\
$D\eta$                 &$-0.5$            &$4.2 $             &$0.9 $             &      &$-1.1  $          &$1.2  $             &$1.3 $                 &&&       $D_s^\ast{K}$           &$-1.9$           &$8.6$            &$2.3 $             &      &$-4.9$            &$1.3  $             &$0.8 $        \\
$D^\ast\eta$            &$-0.3$            &$0.8 $             &$0.2 $             &      &$-1.4  $          &$0.5  $             &$0.4 $                 &&&       $D_0^\ast(2300)\pi$     &$-1.8  $         &$5$E$-3$         &$0.5  $            &      &$-13.5$           &$10.7$              &$11.9$        \\
$D_sK$                  &$-1.3$            &$8.8$              &$0.3 $             &      &$-1.6  $          &$0.9  $             &$0.9 $                 &&&       $D_1(2430)\pi$          &$-4.2  $         &$0.4  $          &$0.5  $            &      &$-0.3  $          &$2.9 $              &$3$E$-3$       \\
$D_s^\ast{K}$           &$-0.6$            &$0.6$              &$0.03$             &      &$-2.7  $          &$0.3  $             &$0.3 $                 &&&       $D_1(2420)\pi$          &$-5.2 $          &$0.7  $          &$0.5$              &      &$-10.6$           &$9.0 $              &$0.8 $        \\
$D_1(2430)\pi$          &$-9.0  $          &$0.02 $            &$0.2  $            &      &$-4.2  $          &$5.6 $              &$0.8 $                 &&&       $D_2^\ast(2460)\pi$     &$-56.5 $         &$66.3$           &$130.3$            &      &$-14.6 $          &$3.8 $              &$6.3 $        \\
$D_1(2420)\pi$          &$-90.1 $          &$62.0$             &$88.9$             &      &$-11.0 $          &$1.1 $              &$0.4 $                 &&&       $D\rho$                 &$-2.1  $         &$2.5 $           &$2.5$              &      &$-0.5  $          &$1.0 $              &$1.0 $        \\
$D_2^\ast(2460)\pi$     &$-12.8 $          &$0.02 $            &$0.01 $            &      &$-17.2 $          &$3.7 $              &$1.1 $                 &&&       $D\omega$               &$-0.7 $          &$0.7 $           &$0.7 $             &      &$-0.2  $          &$0.3$               &$0.3$         \\
$D\rho$                 &$-0.2  $          &$1$E$-5$           &$1$E$-5$           &      &$-0.4  $          &$0.1 $              &$0.1 $                 &&&       \emph{Total}          &$\mathbf{-77.1}$   &$\mathbf{101.2}$ &$\mathbf{162.6}$   &      &$\mathbf{-67.2}$  &$\mathbf{67.1}$    &$\mathbf{53.8}$\\
$D\omega$               &$-0.06 $          &$...$              &$...$              &      &$-0.1  $          &$0.03$              &$0.03$                 &&&       \emph{Exp.}           &$\mathbf{-}$       &\multicolumn{2}{c}{$\mathbf{102\pm32}$} &   &$\mathbf{-}$      &\multicolumn{2}{c}{$\mathbf{-}$}\\
\cline{11-18}
\emph{Total}          &$\mathbf{-119.6}$   &$\mathbf{102.9}$   &$\mathbf{105.1}$   &      &$\mathbf{-58.8}$  &$\mathbf{51.8}$     &$\mathbf{43.9}$        &&&        \multicolumn{8}{c}{\quad}                                                                                                                    \\
\emph{Exp.}           &$\mathbf{-}$        &\multicolumn{2}{c}{$\mathbf{-}$}       &      &$\mathbf{-}$      &\multicolumn{2}{c}{$\mathbf{66\pm5}$}       &&&        \multicolumn{8}{c}{\quad}                                                                                                                        \\
\bottomrule[0.7pt]\bottomrule[0.7pt]
&\multicolumn{3}{c}{$D_s(2^1S_0)$}&
&\multicolumn{3}{c}{$D_s(2^3S_1)$}
&&&
&\multicolumn{3}{c}{$D_s(1^3P_0)$}&
&\multicolumn{3}{c}{$D_s(1^3P_2)$}\\
\cline{2-4}  \cline{6-8} \cline{12-14}  \cline{16-18}
&$\Delta M$   &\multicolumn{2}{c}{$as~D_{s0}(2590)$}&
&$\Delta M$   &\multicolumn{2}{c}{$as~D_{s1}^*(2700)$}
&&&
&$\Delta M$  &\multicolumn{2}{c}{$as~D_{s0}^*(2317)$}&
&$\Delta M$  &\multicolumn{2}{c}{$as~D_{s2}^*(2573)$}\\
\cline{3-4}  \cline{7-8} \cline{13-14}  \cline{17-18}
Channel                &$[2677]$  &$\Gamma_i^{NR}$             &$\Gamma_i^{NR+RC}$&
&$[2753]$  &$\Gamma_i^{NR}$             &$\Gamma_i^{NR+RC}$
&&&Channel
&$[2372]$  &$\Gamma_i^{NR}$             &$\Gamma_i^{NR+RC}$&
&$[2597]$  &$\Gamma_i^{NR}$             &$\Gamma_i^{NR+RC}$\\
\cline{1-4}  \cline{6-8} \cline{11-14}  \cline{16-18}
$DK$                &$...$        &$...$           &$...$          &    &$2.1 $          &$3.7$              &$31.0 $                              &&&    $DK$                  &$-60.0$          &$...$          &$...$                     &     &$-17.9$           &$19.0$                &$16.1 $             \\
$D^\ast{K}$         &$-47.7$      &$17.9$          &$76.2$         &    &$-8.0$          &$3.7$              &$76.8 $                              &&&    $D^\ast K$            &$...$            &$...$          &$...$                     &     &$-20.1$           &$2.2$                 &$1.7 $              \\
$D_s\eta$           &$...$        &$...$           &$...$          &    &$-0.7$          &$0.6$              &$5.8 $                               &&&    $D_s\eta$             &$...$            &$...$          &$...$                     &     &$-2.0$            &$0.1 $                &$0.1  $             \\
$D_s^\ast\eta$      &$...$        &$...$           &$...$          &    &$-3.6$          &$0.1$              &$3.4 $                               &&&    \emph{Total}          &$\mathbf{-60.0}$ &\multicolumn{2}{c}{$\mathbf{-}$}          &     &$\mathbf{-40.0}$  &$\mathbf{21.3}$       &$\mathbf{17.9}$     \\
\emph{Total}        &$\mathbf{-47.7}$  &$\mathbf{17.9}$      &$\mathbf{76.2}$     &    &$\mathbf{-10.2}$&$\mathbf{8.1}$     &$\mathbf{117.0}$      &&&    \emph{Exp.}           &$\mathbf{-}$     &\multicolumn{2}{c}{$\mathbf{<3.8}$}       &     &$\mathbf{-}$      &\multicolumn{2}{c}{$\mathbf{16.9\pm0.7}$} \\
\cline{11-18}
\emph{Exp.}        &$\mathbf{-}$       &\multicolumn{2}{c}{$\mathbf{89\pm20}$}  &    &$\mathbf{-}$    &\multicolumn{2}{c}{$\mathbf{122\pm10}$}                      &&&                           &\multicolumn{3}{c}{$D_s(1P_1)$}                   &     &\multicolumn{3}{c}{$D_s(1P_1)$}                          \\
\cline{1-8}       \cline{12-14}  \cline{16-18}
\multicolumn{8}{c}{\quad}                                                                                                                          &&&                           &$\Delta M$     &\multicolumn{2}{c}{$as~D_{s1}(2460)$}      &     &$\Delta M$  &\multicolumn{2}{c}{$as~D_{s1}(2536)$}             \\
\cline{13-14}  \cline{17-18}
\multicolumn{8}{c}{\quad}                                                                                                                          &&&     Channel            &$[2533]$       &$\Gamma_i^{NR}$ &$\Gamma_i^{NR+RC}$           &     &$[2546]$         &$\Gamma_i^{NR}$  &$\Gamma_i^{NR+RC}$      \\
\cline{11-14} \cline{16-18}
\multicolumn{8}{c}{\quad}                                                                                                                          &&&     $D^\ast K$         &$-49.3$        &$...$     &$...$                    &     &$-28.1$          &$0.6$             &$0.4$                  \\
\multicolumn{8}{c}{\quad}                                                                                                                          &&&     \emph{Total}       &$\mathbf{-49.3}$   &\multicolumn{2}{c}{$\mathbf{-}$}          &     &$\mathbf{-28.1}$ &$\mathbf{0.6}$    &$\mathbf{0.4}$         \\
\multicolumn{8}{c}{\quad}                                                                                                                          &&&     \emph{Exp.}        &$\mathbf{-}$       &\multicolumn{2}{c}{$\mathbf{<3.5}$}       &     &$\mathbf{-}$     &\multicolumn{2}{c}{$\mathbf{0.92\pm0.05}$}  \\
\bottomrule[0.7pt]
&\multicolumn{3}{c}{$D_s(1^3D_1)$}&
&\multicolumn{3}{c}{$D_s(1^3D_3)$}
&&&
&\multicolumn{3}{c}{$D_s(1D_2)$}&
&\multicolumn{3}{c}{$D_s(1D'_2)$}\\
\cline{2-4}  \cline{6-8} \cline{12-14}  \cline{16-18}
&$\Delta M$   &\multicolumn{2}{c}{$as~D_{s1}^*(2860)$}&
&$\Delta M$   &\multicolumn{2}{c}{$as~D_{s3}^*(2860)$}
&&&
&$\Delta M$  &\multicolumn{2}{c}{$M=2869$}&
&$\Delta M$  &\multicolumn{2}{c}{$M=2890$}                             \\
\cline{3-4}  \cline{7-8} \cline{13-14}  \cline{17-18}
Channel                &$[2829]$  &$\Gamma_i^{NR}$   &$\Gamma_i^{NR+RC}$&
&$[2909]$  &$\Gamma_i^{NR}$   &$\Gamma_i^{NR+RC}$
&&&Channel
&$[2877]$  &$\Gamma_i^{NR}$   &$\Gamma_i^{NR+RC}$&
&$[2923]$  &$\Gamma_i^{NR}$   &$\Gamma_i^{NR+RC}$\\
\cline{1-4}  \cline{6-8} \cline{11-14}  \cline{16-18}
$DK$                     &$-1.5 $           &$14.9$              &$13.1$                &    &$-7.3 $          &$16.7$             &$16.0 $             &&&       $D^\ast{K}$                  &$-2.4 $           &$18.7$              &$20.8 $               &     &$-18.2$          &$24.4$               &$22.1 $       \\
$D^\ast{K}$              &$-1.0$            &$7.8$               &$7.0$                 &    &$-10.8$          &$12.0$             &$10.7 $             &&&       $D_s^\ast \eta$              &$-1.5$            &$6.4$               &$2.8$                 &     &$-2.5$           &$1.2$                &$0.8 $        \\
$D_s\eta$                &$-1.2 $           &$7.3$               &$1.6$                 &    &$-0.8 $          &$0.9$              &$0.6 $              &&&       $D_0^\ast (2300)K$           &$-1.5 $           &$2$E$-6$            &$4$E$-4$              &     &$-11.5$          &$0.8$                &$0.9 $              \\
$D_s^\ast \eta$          &$-0.6 $           &$2.4$               &$0.6$                 &    &$-1.5 $          &$0.4$              &$0.3 $              &&&       $DK^\ast$                    &$-2.3$            &$2.9$               &$2.8$                 &     &$-0.5$           &$0.9$                &$1.0 $              \\
$DK^\ast$                &$-0.1$            &$0.04$              &$0.03$                &    &$-0.4$           &$0.1$              &$0.1 $              &&&       \emph{Total}                 &$\mathbf{-7.7}$   &$\mathbf{28.0}$     &$\mathbf{26.4}$       &     &$\mathbf{-32.7}$ &$\mathbf{27.3}$      &$\mathbf{24.8}$     \\
\emph{Total}             &$\mathbf{-4.4}$   &$\mathbf{32.4}$     &$\mathbf{22.3}$       &    &$\mathbf{-20.8}$     &$\mathbf{30.1}$    &$\mathbf{27.7}$     &&&       \emph{Exp.}                  &$\mathbf{-}$  &\multicolumn{2}{c}{$\mathbf{-}$}           &     &$\mathbf{-}$     &\multicolumn{2}{c}{$\mathbf{-}$}   \\
\cline{11-18}
\emph{Exp.}            &$\mathbf{-}$      &\multicolumn{2}{c}{$\mathbf{159\pm122}$} &    &$\mathbf{-}$     &\multicolumn{2}{c}{$\mathbf{53\pm17}$} &&&        \multicolumn{8}{c}{\quad}       \\
\bottomrule[1.0pt]\bottomrule[1.0pt]
\end{tabular*}}
\end{table*}

\begin{table*}[htbp]
\caption{The mass shifts $\Delta M$ and partial widths (in MeV) of the $B$ and $B_s$ states.
	The bare masses obtained from the potential model are listed in square brackets. The strong decay widths, which combine the unquenched spectra and are described by non-relativistic chiral interactions $\mathcal{H}_{I}^{NR}$, are further augmented by a relativistic correction term $\mathcal{H}_{I}^{RC}$. These widths are denoted as $\Gamma_i^{NR}$ and $\Gamma_i^{NR+RC}$, respectively. The forbidden decay channel is denoted by ``$...$". The experimental data are taken from the PDG~\cite{ParticleDataGroup:2022pth}.} \label{BBsmeson}
\renewcommand\arraystretch{1.10}
\tabcolsep=0.154cm
\resizebox{\textwidth}{!}{
\begin{tabular*}{190mm}{ccccccccc|lccccccccc}
\bottomrule[1.0pt]\bottomrule[1.0pt]
                       &\multicolumn{3}{c}{$B(2^1S_0)$}&
&\multicolumn{3}{c}{$B(2^3S_1)$}
&&&
&\multicolumn{3}{c}{$B(1^3P_0)$}&
&\multicolumn{3}{c}{$B(1^3P_2)$}\\ \cline{2-4}  \cline{6-8} \cline{12-14}  \cline{16-18}
&$\Delta M$   &\multicolumn{2}{c}{$M=5789$}&
&$\Delta M$   &\multicolumn{2}{c}{$as~B_J(5840)^+$}
&&&
&$\Delta M$  &\multicolumn{2}{c}{$M=5616$}&
&$\Delta M$  &\multicolumn{2}{c}{$as~B_2^*(5747)^{+/0}$}\\
\cline{3-4}  \cline{7-8} \cline{13-14}  \cline{17-18}
Channel                &$[5882]$    &$\Gamma_i^{NR}$   &$\Gamma_i^{NR+RC}$&
&$[5908]$    &$\Gamma_i^{NR}$   &$\Gamma_i^{NR+RC}$
&&&Channel
&$[5704]$    &$\Gamma_i^{NR}$   &$\Gamma_i^{NR+RC}$&
&$[5780]$    &$\Gamma_i^{NR}$   &$\Gamma_i^{NR+RC}$\\
\cline{1-4}  \cline{6-8} \cline{11-14}  \cline{16-18}
$B\pi$                &$...$             &$...$            &$...$         &    &$-2.5$            &$0.2$             &$95.9 $                                    &&& $B\pi$                &$-88.4$              &$138.3$                 &$298.7$          &     &$-19.5$           &$21.9/22.4$             &$15.1/15.5$                    \\
$B^\ast\pi$           &$-34.8$               &$18.8$               &$256.7$           &    &$-13.2$           &$3.7 $            &$175.5$                                    &&& $B^\ast\pi$           &$...$            &$...$               &$...$        &     &$-26.0$           &$20.5/21.1$             &$13.1/13.6$                    \\
$B\eta$               &$...$             &$...$            &$...$         &    &$-1.4$            &$0.2 $            &$1.0  $                                    &&& \emph{Total}          &$\mathbf{-88.4}$     &$\mathbf{138.3}$        &$\mathbf{298.7}$ &     &$\mathbf{-45.5}$  &$\mathbf{42.4/43.5}$    &$\mathbf{28.2/29.1}$       \\
$B^\ast\eta$          &$...$             &$...$            &$...$         &    &$-2.1$            &$...$         &$...$                                  &&&  \emph{Exp.}          &$\mathbf{-}$         &\multicolumn{2}{c}{$\mathbf{-}$}          &     &$\mathbf{-}$     &\multicolumn{2}{c}{$\mathbf{24.2\pm1.7/20\pm5}$}   \\
\cline{11-18}
$B_sK$                &$...$             &$...$            &$...$         &    &$-2.1$            &$...$         &$...$                                  &&&  \multicolumn{8}{c}{\quad}  \\
$B_s^\ast{K}$         &$...$             &$...$            &$...$         &    &$-3.5 $           &$...$         &$...$                                  &&&  \multicolumn{8}{c}{\quad}  \\
\cline{11-18}
$B(1^3P_0)\pi$        &$-58.2$               &$17.1$               &$18.9$            &    &$...$         &$...$         &$...$                                  &&&  &\multicolumn{3}{c}{$B(1P_1)$}                                                        &&\multicolumn{3}{c}{$B(1P'_1)$}     \\
\cline{12-14}  \cline{16-18}
$B(1P_1)\pi$             &$...$          &$...$            &$...$         &    &$-44.9$           &$7.9$             &$9.2 $                                     &&&  &$\Delta M$&\multicolumn{2}{c}{$M=5691$}                                              &&$\Delta M$&\multicolumn{2}{c}{$as~B_1(5721)^{+/0}$}     \\
\cline{13-14}  \cline{17-18}
$B_1(5721)\pi$           &$...$          &$...$            &$...$         &    &$-6.1 $           &$...$         &$...$                                  &&&  Channel            &$[5757]$           &$\Gamma_i^{NR}$      &$\Gamma_i^{NR+RC}$      &&$[5771]$            &$\Gamma_i^{NR}$         &$\Gamma_i^{NR+RC}$     \\
\cline{11-18}
$B_2^\ast(5747)\pi$      &$...$          &$...$            &$...$         &    &$-10.0$           &$...$         &$...$                                  &&&  $B^\ast\pi$        &$-66.2$            &$126.5$              &$310.7$                 &&$-41.3$             &$29.0/29.0$             &$16.5/16.5$              \\
\emph{Total}             &$\mathbf{-93.0}$   &$\mathbf{35.9}$      &$\mathbf{275.6}$  &    &$\mathbf{-85.8}$  &$\mathbf{12.0}$   &$\mathbf{281.6}$                           &&&  \emph{Total}       &$\mathbf{-66.2}$   &$\mathbf{126.5}$     &$\mathbf{310.7}$        &&$\mathbf{-41.3}$    &$\mathbf{29.0/29.0}$    &$\mathbf{16.5/16.5}$   \\
\emph{Exp.}              &$\mathbf{-}$       &\multicolumn{2}{c}{$\mathbf{-}$}        &    &$\mathbf{-}$      &\multicolumn{2}{c}{$\mathbf{224\pm104}$}                      &&&  \emph{Exp.}        &$\mathbf{-}$       &\multicolumn{2}{c}{$\mathbf{-}$}              &&$\mathbf{-}$        &\multicolumn{2}{c}{$\mathbf{27.5\pm3.4/31\pm6}$} \\
\bottomrule[0.7pt]
&\multicolumn{3}{c}{$B(1^3D_1)$}&
&\multicolumn{3}{c}{$B(1^3D_3)$}
&&&
&\multicolumn{3}{c}{$B(1D_2)$}&
&\multicolumn{3}{c}{$B(1D'_2)$}\\ \cline{2-4}  \cline{6-8} \cline{12-14}  \cline{16-18}
&$\Delta M$   &\multicolumn{2}{c}{$M=5951$}&
&$\Delta M$   &\multicolumn{2}{c}{$M=6004$}
&&&
&$\Delta M$  &\multicolumn{2}{c}{$M=5955$}&
&$\Delta M$  &\multicolumn{2}{c}{$M=6049$}\\
\cline{3-4}  \cline{7-8} \cline{13-14}  \cline{17-18}
Channel                &$[6095]$    &$\Gamma_i^{NR}$   &$\Gamma_i^{NR+RC}$&
&$[6070]$    &$\Gamma_i^{NR}$   &$\Gamma_i^{NR+RC}$
&&&Channel
&$[6055]$    &$\Gamma_i^{NR}$   &$\Gamma_i^{NR+RC}$&
&$[6129]$    &$\Gamma_i^{NR}$   &$\Gamma_i^{NR+RC}$\\
\cline{1-4}  \cline{6-8} \cline{11-14}  \cline{16-18}
$B\pi$                  &$-4.6$          &$24.1$           &$18.2$       &        &$-11.8  $      &$22.7 $                &$22.8$                &&&       $B^\ast\pi$             &$-6.5$           &$29.6$                &$25.3$               &     &$-28.5 $     &$55.4 $              &$52.8$                     \\
$B^\ast\pi$             &$-2.4$          &$11.7$           &$7.9 $       &        &$-15.3 $       &$23.9 $                &$22.8$                &&&       $B^\ast\eta$            &$-1.0$           &$3.9$                 &$0.8$                &     &$-3.3  $     &$1.4  $              &$1.1 $                     \\
$B\eta$                 &$-0.7$          &$4.4 $           &$0.9 $       &        &$-1.4  $       &$0.5  $                &$0.5 $                &&&       $B_s^\ast{K}$           &$-2.0$           &$3.5$                 &$0.2$                &     &$-6.1$       &$1.3  $              &$0.7 $                     \\
$B^\ast\eta$            &$-0.3$          &$1.3 $           &$0.2 $       &        &$-1.6  $       &$0.3  $                &$0.2 $                &&&       $B(1^3P_0)\pi$          &$-2.9  $         &$1$E$-4$              &$0.3 $               &     &$-14.7 $     &$14.7 $              &$18.8 $                    \\
$B_s K$                 &$-1.6$          &$7.0$            &$0.1 $       &        &$-2.0  $       &$0.2  $                &$0.2 $                &&&       $B(1P_1)\pi$            &$-3.8 $          &$0.2  $               &$0.3$                &     &$-0.6 $      &$3.7 $               &$0.4 $                     \\
$B_s^\ast K$            &$-0.8  $        &$1.1  $          &$0.02  $     &        &$-3.3  $       &$0.1 $                 &$0.1 $                &&&       $B_1(5721)\pi$          &$-4.7$           &$0.09$                &$0.1$                &     &$-9.1 $      &$6.5$                &$0.1 $                     \\
$B(1P_1)\pi$            &$-11.2 $        &$0.1  $          &$1.7  $      &        &$-2.1 $        &$5.5 $                 &$0.2 $                &&&       $B_2^*(5747)\pi$        &$-79.5 $         &$51.9$                &$78.3$               &     &$-17.8 $     &$4.6 $               &$8.7 $                     \\
$B_1(5721)\pi$          &$-105.8$        &$81.3$           &$123.2$      &        &$-12.9 $       &$0.5 $                 &$0.1 $                &&&       \emph{Total}           &$\mathbf{-100.4}$ &$\mathbf{89.2}$       &$\mathbf{105.3}$     &&$\mathbf{-80.1}$   &$\mathbf{87.6}$    &$\mathbf{82.6}$   \\
\cline{11-18}
$B_2^\ast (5747)\pi$    &$-16.7 $        &$0.1$            &$0.1$        &        &$-15.4 $       &$1.9 $                 &$0.4 $                &&&       \multicolumn{8}{c}{\quad}                                                                                                                \\
\emph{Total}          &$\mathbf{-144.1}$ &$\mathbf{131.1}$ &$\mathbf{152.2}$   &&$\mathbf{-65.8}$  &$\mathbf{55.6}$    &$\mathbf{47.3}$          &&&       \multicolumn{8}{c}{\quad}                                                                                                                \\
\bottomrule[0.7pt]\bottomrule[0.7pt]
&\multicolumn{3}{c}{$B_s(2^1S_0)$}&
&\multicolumn{3}{c}{$B_s(2^3S_1)$}
&&&
&\multicolumn{3}{c}{$B_s(1^3P_0)$}&
&\multicolumn{3}{c}{$B_s(1^3P_2)$}\\ \cline{2-4}  \cline{6-8} \cline{12-14}  \cline{16-18}
&$\Delta M$   &\multicolumn{2}{c}{$M=5915$}&
&$\Delta M$   &\multicolumn{2}{c}{$M=5953$}
&&&
&$\Delta M$  &\multicolumn{2}{c}{$M=5711$}&
&$\Delta M$  &\multicolumn{2}{c}{$as~B_{s2}^*(5840)^{0}$}\\
\cline{3-4}  \cline{7-8} \cline{13-14}  \cline{17-18}
Channel                &$[5969]$    &$\Gamma_i^{NR}$   &$\Gamma_i^{NR+RC}$&
&$[5996]$    &$\Gamma_i^{NR}$   &$\Gamma_i^{NR+RC}$
&&&Channel
&$[5772]$    &$\Gamma_i^{NR}$   &$\Gamma_i^{NR+RC}$&
&$[5864]$    &$\Gamma_i^{NR}$   &$\Gamma_i^{NR+RC}$\\
\cline{1-4}  \cline{6-8} \cline{11-14}  \cline{16-18}
$BK$                 &$...$        &$...$           &$...$         &    &$-9.4$      &$7.4$       &$70.1 $                            &&&    $BK$                &$-60.9$           &$...$           &$...$           &     &$-16.1$           &$2.56$                 &$1.91 $             \\
$B^\ast{K}$          &$-53.6$          &$34.3$              &$131.1$           &    &$-25.8$     &$19.3$      &$108.0$                            &&&    $B^\ast K$          &$...$         &$...$           &$...$           &     &$-20.7$           &$0.23$                 &$0.16 $             \\
$B_s\eta$            &$...$        &$...$           &$...$         &    &$-3.4 $     &$0.05$      &$2.6 $                             &&&    \emph{Total}        &$\mathbf{-60.9}$  &\multicolumn{2}{c}{$\mathbf{-}$}           &     &$\mathbf{-36.8}$  &$\mathbf{2.79}$        &$\mathbf{2.07}$            \\
$B_s^*\eta$          &$...$        &$...$           &$...$         &    &$-4.7 $     &$...$   &$...$                          &&&    \emph{Exp.}         &$\mathbf{-}$      &\multicolumn{2}{c}{$\mathbf{-}$}         &     &$\mathbf{-}$      &\multicolumn{2}{c}{$\mathbf{1.49\pm0.27}$}    \\
\cline{11-18}
\emph{Total}        &$\mathbf{-53.6}$  &$\mathbf{34.3}$     &$\mathbf{131.1}$  && $\mathbf{-43.3}$ &$\mathbf{26.8}$    &$\mathbf{180.7}$         &&&     &&\multicolumn{3}{c}{$B_s(1P_1)$}   &&\multicolumn{3}{c}{$B_s(1P'_1)$}        \\
\cline{12-14}  \cline{16-18}
\emph{Exp.}       &$\mathbf{-}$       &\multicolumn{2}{c}{$\mathbf{-}$}          &    &$\mathbf{-}$ &\multicolumn{2}{c}{$\mathbf{-}$}            &&&     &$\Delta M$         &\multicolumn{2}{c}{$M=5781$}  &&$\Delta M$&\multicolumn{2}{c}{$as~B_{s1}(5830)^0$}       \\
\cline{1-8}\cline{13-14}  \cline{17-18}
\multicolumn{8}{c}{\quad}                                                                                                                        &&&     Channel             &$[5840]$           &$\Gamma_i^{NR}$        &$\Gamma_i^{NR+RC}$            &     &$[5842]$        &$\Gamma_i^{NR}$   &$\Gamma_i^{NR+RC}$         \\
\cline{11-18}
\multicolumn{8}{c}{\quad}                                                                                                                        &&&     $B^\ast K$          &$-59.3$            &$...$              &$...$                     &    &$-31.5$          &$0.06$            &$0.04$                                                  \\
\multicolumn{8}{c}{\quad}                                                                                                                        &&&     \emph{Total}        &$\mathbf{-59.3}$   &\multicolumn{2}{c}{$\mathbf{-}$}                      &    &$\mathbf{-31.5}$ &$\mathbf{0.06}$   &$\mathbf{0.04}$             \\
\multicolumn{8}{c}{\quad}                                                                                                                        &&&     \emph{Exp.}         &$\mathbf{-}$       &\multicolumn{2}{c}{$\mathbf{-}$}                      &    &$\mathbf{-}$     &\multicolumn{2}{c}{$\mathbf{0.5\pm0.4}$} \\
\bottomrule[0.7pt]
&\multicolumn{3}{c}{$B_s(1^3D_1)$}&
&\multicolumn{3}{c}{$B_s(1^3D_3)$}
&&&
&\multicolumn{3}{c}{$B_s(1D_2)$}&
&\multicolumn{3}{c}{$B_s(1D'_2)$}\\ \cline{2-4}  \cline{6-8} \cline{12-14}  \cline{16-18}
&$\Delta M$   &\multicolumn{2}{c}{$M=6149$}&
&$\Delta M$   &\multicolumn{2}{c}{$M=6116$}
&&&
&$\Delta M$  &\multicolumn{2}{c}{$as~B_{sJ}(6109)$}&
&$\Delta M$  &\multicolumn{2}{c}{$as~B_{sJ}(6158)$}\\
\cline{3-4}  \cline{7-8} \cline{13-14}  \cline{17-18}
Channel                &$[6156]$    &$\Gamma_i^{NR}$   &$\Gamma_i^{NR+RC}$&
&$[6146]$    &$\Gamma_i^{NR}$   &$\Gamma_i^{NR+RC}$
&&&Channel
&$[6131]$    &$\Gamma_i^{NR}$   &$\Gamma_i^{NR+RC}$&
&$[6191]$    &$\Gamma_i^{NR}$   &$\Gamma_i^{NR+RC}$\\
\cline{1-4}  \cline{6-8} \cline{11-14}  \cline{16-18}
$BK$                &$-2.7 $           &$28.3$           &$24.3$           &    &$-11.8 $         &$17.6$               &$17.9$               &&&       $B^\ast{K}$         &$-4.2 $           &$41.4$           &$29.9 $                   &    &$-28.2$           &$41.7$              &$40.5 $           \\
$B^\ast{K}$         &$-1.7$            &$15.2$           &$11.7$           &    &$-15.0$          &$16.4$               &$16.0$               &&&       $B_s^*\eta$         &$-2.2 $           &$10.5$           &$2.2$                     &    &$-3.6$            &$1.7$               &$1.2 $            \\
$B_s \eta$          &$-1.6 $           &$12.9$           &$2.5$            &    &$-1.2 $          &$0.4$                &$0.4 $               &&&       \emph{Total}        &$\mathbf{-6.4}$   &$\mathbf{51.9}$  &$\mathbf{32.1}$           &    &$\mathbf{-31.8}$  &$\mathbf{43.4}$    &$\mathbf{41.7}$   \\
$B_s^\ast \eta$     &$-0.8$            &$5.3$            &$1.0$            &    &$-2.0$           &$0.4$                &$0.3 $               &&&       \emph{Exp.}        &$\mathbf{-}$       &\multicolumn{2}{c}{$\mathbf{22\pm9}$}       &   &$\mathbf{-}$   &\multicolumn{2}{c}{$\mathbf{72\pm43}$}   \\
\cline{11-18}
\emph{Total}      &$\mathbf{-6.8}$   &$\mathbf{61.7}$  &$\mathbf{39.5}$  &    &$\mathbf{-30.0}$ &$\mathbf{34.8}$     &$\mathbf{34.6}$      &&&       \multicolumn{8}{c}{\quad}                                                                                                                  \\
\bottomrule[1.0pt]\bottomrule[1.0pt]

\end{tabular*}}
\end{table*}

\end{document}